\documentclass[modern]{aastex62}

\usepackage{gensymb}
\usepackage{amsmath}
\usepackage{graphicx}
\usepackage{booktabs}
\usepackage{threeparttable}
\usepackage{multirow}
\usepackage{float}
\usepackage[graphicx]{realboxes}
\usepackage{rotating}
\usepackage[normalem]{ulem}

\definecolor{orange}{RGB}{255,127,0}

\newcommand{\crpf}[1]{\textcolor{cyan}{particle $s^{-1} cm^{-2}$}}

\graphicspath{{./}{figures/}}

\received{1 October 2020}
\revised{19 April 2021}
\accepted{20 April 2021}

\submitjournal{ApJ}

\shorttitle{HST Cosmic Rays}
\shortauthors{Miles, Deustua, Tancredi et al}

\begin{document}

\title{Using Cosmic Rays detected by HST as Geophysical Markers I:  Detection and Characterization of Cosmic Rays}

\author[0000-0001-8936-4545]{Nathan D. Miles}
\affiliation{Space Telescope Science Institute, 
3700 San Martin Drive,
Baltimore, MD 21218, USA}
\affiliation{Department of Earth, Planetary, and Space Sciences, University of California Los Angeles, Los Angeles, CA, USA}
\author[0000-0003-2823-360X]{Susana E. Deustua}
\affiliation{Space Telescope Science Institute,
3700 San Martin Drive, 
Baltimore, MD 21218, USA}
\author[0000-0002-4943-8623]{Gonzalo Tancredi}
\affiliation{Universidad de la Rep\'ublica, 
Montevideo, Uruguay}
\author[0000-0001-9482-0410]{Germ\'an Schnyder}
\affiliation{Universidad de la Rep\'ublica, 
Montevideo, Uruguay}
\author[0000-0002-8146-4012]{Sergio Nesmachnow}
\affiliation{Universidad de la Rep\'ublica, 
Montevideo, Uruguay}
\author[0000-0001-8481-405X]{Geoffrey Cromwell}
\affiliation{Current Address: U.S. Geological Survey, California Water Science Center, Santa Maria, CA}
\nocollaboration

\correspondingauthor{Nathan Miles and Susana E. Deustua}
\email{ndmiles@ucla.edu,  deustua@arsmetrologia.org}

\begin{abstract}
The Hubble Space Telescope (HST) has been operational for over 30 years and throughout that time it has been bombarded by high energy charged particles colloquially referred to as cosmic rays. In this paper, we present a comprehensive study of more than 1.2 billion cosmic rays observed with HST using a custom written python package, \texttt{HSTcosmicrays}, that is available to the astronomical community. We analyzed $75,908$ dark calibration files taken as part of routine calibration programs for five different CCD imagers with operational coverage of Solar Cycle 23 and 24. We observe the expected modulation of galactic cosmic rays by solar activity. We model the observed energy-loss distributions to derive an estimate of 534 $\pm$ 117 MeV for the kinetic energy of the typical cosmic ray impacting HST. For the three imagers with the largest non-uniformity in thickness, we independently confirm the overall structure produced by fringing analyses by analyzing cosmic ray strikes across the detector field of view. We analyze STIS/CCD observations taken as HST crosses over the South Atlantic Anomaly and find a peak cosmic-ray particle flux of $\sim1100$ $particle/s/cm^2$. We find strong evidence for two spatially confined regions over North America and Australia that exhibit increased cosmic-ray particle fluxes at the $5\sigma$ level.

\end{abstract}


\section{Introduction}\label{sec:intro}
\begin{figure}[H]
    \centering
    \includegraphics[width=\textwidth]{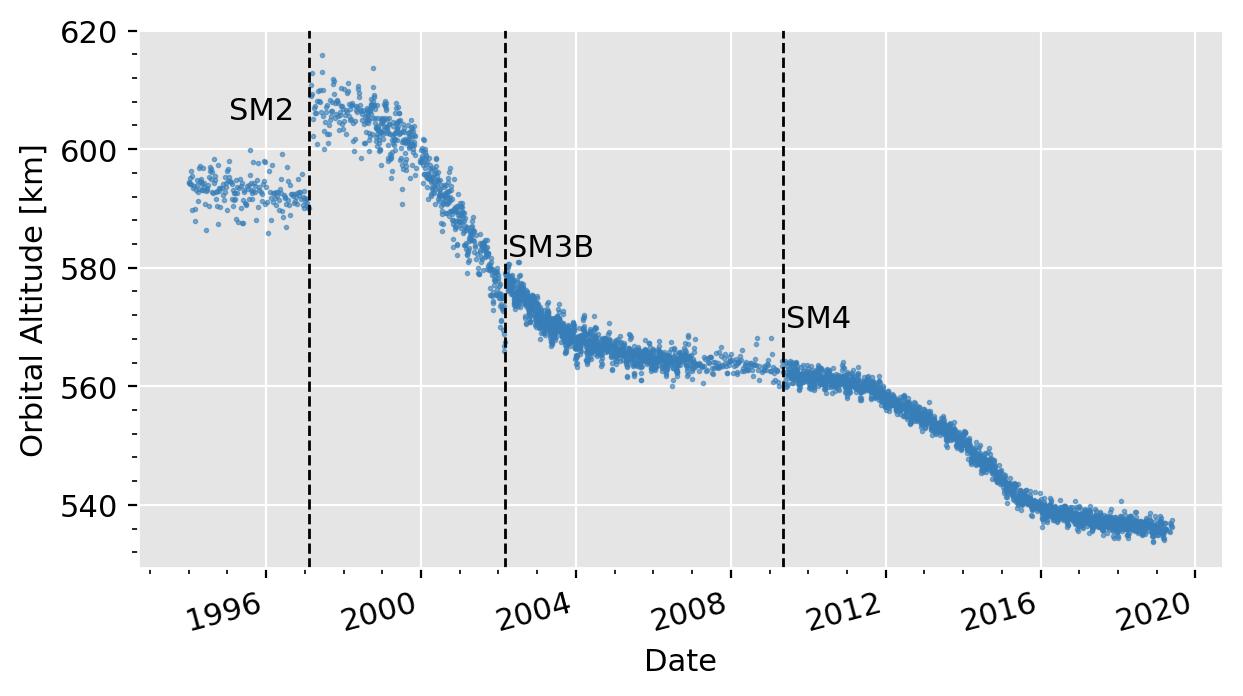}
    \caption{The orbital altitude of HST as a function of time computed from telemetry data. Vertical lines indicate servicing missions where instruments were added and/or removed. The observatory received two re-boosts, one in Servicing Mission 2 (SM2) and one in Servicing Mission 3B (SM3B) to compensate for orbital decay. The periods of accelerated orbital decay are driven by contemporaneous solar maxima in Solar Cycles 23 and 24. The increased solar activity increases the density in the thermosphere, which extends from ~60km to ~750km and encompassess the orbit of HST. This leads to an increase in satellite drag and an accelerated orbital decay \citep{walterscheid1989}.}
    \label{fig:altitude}
\end{figure}

The Hubble Space Telescope's (HST) four generations of instruments have enabled scientific research since 1990, providing key data for new astrophysical discoveries. Operating at its current orbital altitude of roughly 538 km (Figure \ref{fig:altitude}) above the Earth's surface, HST is not shielded by the terrestrial atmosphere and so every image obtained with a solid state detector is polluted with charged particle events. These charged particles originate in the solar wind, coronal mass ejections, and elsewhere in the Milky Way galaxy as a result of energetic astrophysical processes (e.g. supernovae, accretion driven phenomena like jets). Observers design their programs to minimize the effect of cosmic rays on astronomical images, and significant effort goes into developing software to identify, flag, and remove these cosmic rays so that the acquired data are useful for astronomical science analysis. However, this process also throws away information that could be used for geophysical investigations.  

For example, the incident flux of charged particles at the Earth is affected by the strength, stability, and morphology of the Earth's magnetosphere. Upon entering the Earth's magnetosphere, these particles are immediately subjected to the Lorentz forces associated with their motion through the Earth's magnetic field (hereafter referred to as the geomagnetic field). Low energy particles become bound to the field lines and follow helical trajectories towards the poles, while the high energy particle penetrate deeper into the upper atmosphere. The nature of this interaction provides us with an opportunity to examine the geomagnetic field in the orbital environment of HST by extracting and analyzing cosmic rays in HST observations. 

The geogmagnetic field is described by a dipole with higher order terms that reflect short term variations. These short term variations occur on timescales of seconds to years and their contributions produce non-uniform departures in direction and intensity from the dipole field \citep{igrf2015}. The most well-known departure is a region known as the South Atlantic Anomaly (SAA), where the local magnetic field measured at the Earth's surface is significantly weaker \citep{schaefer2016}. The decreased magnetic field intensity of the SAA reduces shielding from high energy charged particles. This increases the amount of radiation damage to spacecraft in low earth orbit (LEO) which includes the International Space Station (ISS) and its human inhabitants. The orbital trajectory of HST regularly crosses the SAA and observations along the boundary are used to help map the extent of SAA at an altitude of $\sim540$ km, assess its effects on observations, and to provide a means of monitoring the efficacy of the SAA avoidance contours used in the scheduling of HST observations \citep{lupie2002, barker2009, martel2009}.

Over the course of the 11 year Solar Cycle, fluctuations in solar activity modulate the flux of galactic cosmic rays (GCRs) \citep{potgieter2013}; increased solar activity results in more scattering and decreases the GCR flux at Earth and vice versa. The first evidence of this modulation was observed in data obtained with ion chambers at four stations \citep{Forbush_1954} and then by a neutron monitoring system that was established in 1958 for the International Geophysical Year \citep{Simpson_1958}. The neutron monitoring system provided continuous and standardized measurements of the GCR flux and confirmed the modulation \citep{Lockwood_1958}. By the 1990s, the 11-yr cycle and the interplanetary origin of the GCR intensity variation was firmly established when data were obtained from satellites and space probes \citep{Simpson_1994}. By analyzing charged particle rates in HST observations, we examine the effects of solar activity on the overall incident particle flux as a function of time in the orbital environment of HST.

As described in \citet{Schnyder2017_a} and \citet{Schnyder2017_b}, we deployed a cloud based distributed computing platform for processing HST observations to identify and analyze comsic rays. This early implementation was based on python wrapped around the IRAF command language, and served as a proof of concept. However, the IRAF command language is not designed for batch processing of large datasets and is currently no longer supported. Because of this we developed the open-source Python package, \texttt{HSTCosmicrays}, to identify and analyze cosmic rays in HST images.   

In this paper, we describe our software and highlight key results that demonstrate our ability to reliably identify and analyze cosmic rays. In Section \ref{s:dataset}, we describe the $\sim$4.6TB of HST calibration data used in the analysis. In Section \ref{s:hstcosmicrays}, we provide an overview of the pipeline, describe the cosmic ray rejection algorithms, and list the data extracted. In Section \ref{s:results}, we present the results of analyzing more than 1.2 billion cosmic rays observed over a period of $\sim$25 years which includes the following; cosmic ray morphology in Section \ref{s:morphology}, cosmic ray track lengths in Section \ref{s:path_lengths}, the observed cosmic-ray particle fluxes in Section \ref{s:cr_fluxes}, cosmic rays as a proxy for detector thickness in Section \ref{s:thickness}, modulation by the solar cycle in Section \ref{s:solar_modulation}, cosmic rays in the SAA in Section \ref{s:saa}, spatially correlated cosmic ray "hot spots" in Section \ref{s:hot_spots}, and an estimation of the observed cosmic rays average kinetic energy in Section \ref{s:energy_estimation}. A detailed geophysical analysis is beyond the scope of this paper.

\section{Dataset}\label{s:dataset}

In this work, we analyzed images taken with five different CCD imagers on four instruments: Wide Field Planetary Camera 2 (WFPC2), Space Telescope Imaging Spectrograph (STIS), the High Resolution Channel (HRC) and Wide Field Channel (WFC) in the  Advanced Camera for Surveys (ACS) , and, the  Wide Field Camera 3 (WFC3) UVIS channel. In Table \ref{tab: detectors} we list the detector characteristics relevant to our analysis. 
\begin{table}[h]
\caption{Properties of the CCD imagers analyzed. For CCDs with multiple chips (i.e. ACS/WFC, WFPC2, and WFC3/UVIS), the detector size is the combined area of all the chips. For the detector type, "F" corresponds to thick, frontside-illuminated CCDs and "B" corresponds to thin, backside illuminated CCDs.}

\begin{tabular}{ccccc}
\toprule
Instrument & Detector  & Epitaxial Layer   & Operational   & Detector \\
 &  Size ($cm^2$) &  Thickness ($\mu m$)  &  Period & Type \\
\midrule
WFPC2 & 5.76 & $ \sim10 $ & 01/1994 - 05/2009 & F\\
STIS/CCD & 4.624 & $13.24 - 14.83$ & 02/1997 - 08/2004, 05/2009 - & B \\
ACS/HRC & 4.624 & $12.49 - 16.03$ & 03/2002 - 01/2007 & B\\
ACS/WFC & 37.748 & $12.60 - 17.10$ & 03/2002 - 01/2007, 05/2009 -  & B \\
WFC3/UVIS & 37.804 &$ 13.50 - 18.00 $ & 05/2009 -  & B \\
\bottomrule
\end{tabular}
\label{tab: detectors}
\end{table}

\begin{figure}[h]
    \centering
    \includegraphics[width=\textwidth]{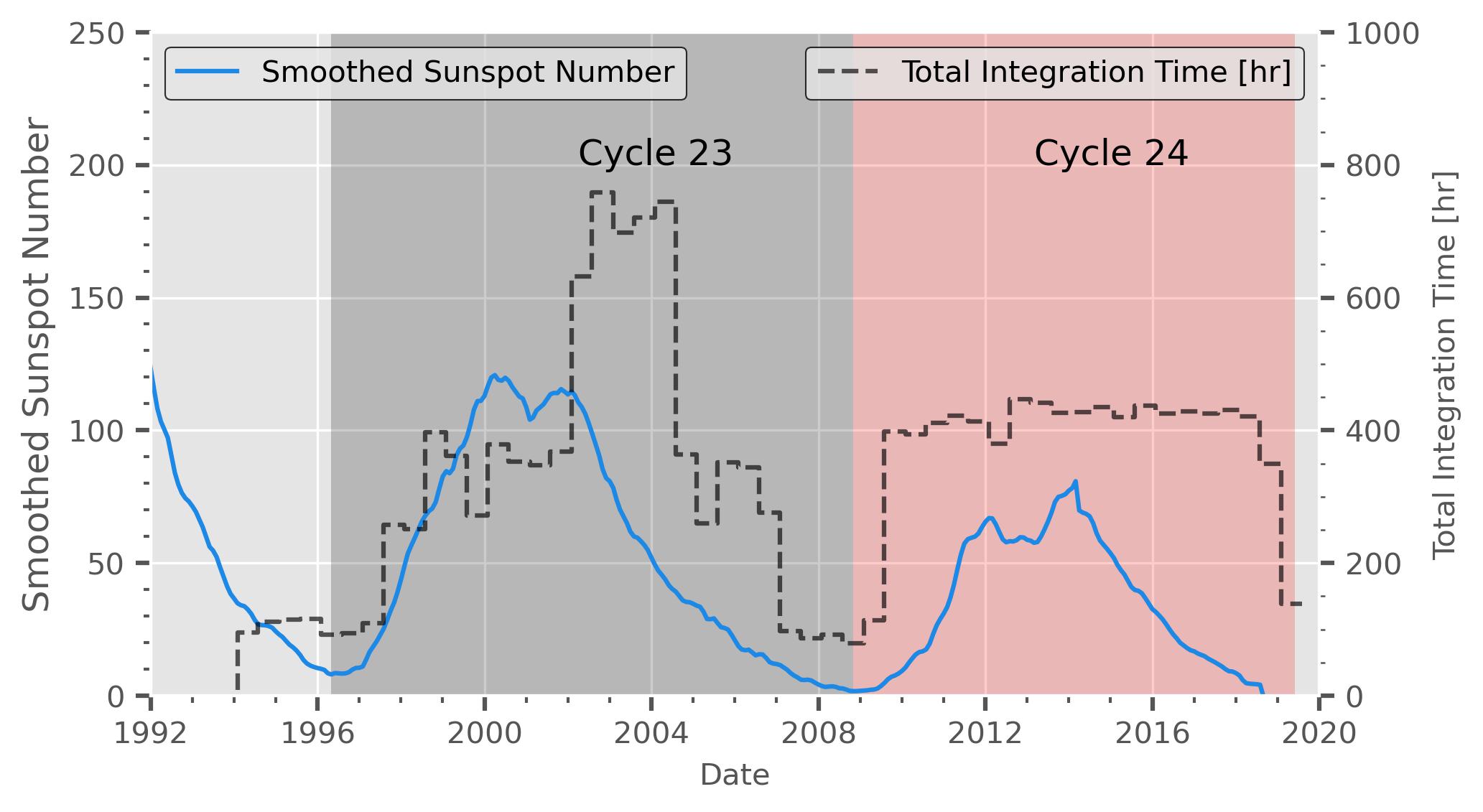}
    \caption{The 13-month smoothed, monthly sunspot number reported by the the WDC-SILSO, Royal Observatory of Belgium, Brussels \citep{sidc}. The gray shading indicates Solar Cycle 23, and the red shading indicates Solar Cycle 24. The dashed, black line is a histogram of the integration times for all instruments in 6 month bins. On average, the dataset contains roughly $\sim$1.8 hours of HST observations every day from 1994 to 2019.5 with complete coverage of Solar Cycles 23 and 24. The significant drop starting in 2005 is due to the STIS failure in August of 2004 and the ACS failure in January of 2007 (see Table \ref{tab: detectors}).}
    \label{fig: solar_cycle}
\end{figure}

The detector size and thickness determine the overall cross section for interaction between cosmic rays and the detection layer of the CCD substrate and this allows us to probe the detector properties. The combined period of operation of the five imagers provides continuous coverage of Solar Cycle 23 and Solar Cycle 24 (Figure \ref{fig: solar_cycle}) allowing us to probe the effects of solar activity on the cosmic ray flux at HST's orbital altitude. 

We restrict our analysis to dark calibration frames (hereafter referred to as darks). Darks are images taken with the shutter closed and they are used to quantify and remove the thermal noise (or dark current) present in the CCDs \citep{janesick2001}. As part of the standard calibration, each dark is gain-calibrated to convert from units of DN to units of electrons and has been bias-corrected and dark subtracted. Since the shutter is closed charged particles are the only external sources present. Thus any signal above the background noise level is due to the interaction between charged particles and the silicon atoms in the epitaxial layer. This facilitates the identification process by completely eliminating any chance of confusion with transient astrophysical sources. Additionally, the entire dataset of darks is available in the cloud as part of the \href{https://registry.opendata.aws/hst/}{HST Public Dataset} hosted on Amazon Web Services (AWS). This gives us the ability to leverage the compute resources and network infrastructure of AWS to boost the performance of our software. 

\begin{table}[H]
\centering
\caption{The dataset for each imager}
\begin{tabular}{cccc}
\toprule
Instrument & Image Count & Data Volume (TB) & Total EXPTIME (hr)  \\
\midrule
WFPC2      & 13,317      & 0.131            & 5,098                    \\
STIS/CCD   & 31,430      & 0.311            & 3,765                     \\
ACS/HRC    & 5,477       & 0.055            & 1,462                     \\
ACS/WFC    & 13,311      & 2.130            & 3,498                    \\
WFC3/UVIS  & 12,373      & 1.980            & 3,040                    \\
\midrule
Totals     & 75,908      & 4.607            & 16,863               \\
\bottomrule
\end{tabular} 
\label{tab: dataset}
\end{table}

\section{\texttt{HSTcosmicrays}}\label{s:hstcosmicrays}

\subsection{Pipeline Overview}\label{s:pipeline_overview}

\texttt{HSTcosmicrays} is written entirely in python and is available on Github\footnote{\url{https://github.com/nmiles2718/hst\_cosmic\_rays}}.  We optimize runtime with \texttt{dask} \citep{dask}, a Python parallelization framework. We store the results for each dataset in HDF5 \citep{hdf5} format and the package contains a module for reading/writing of data. In order to handle the $4.6$TB data volume associated with all $75,908$ images, the pipeline was designed to be lightweight and modular and the benefits of this are two fold. 

First, the storage requirements are now tied to the data generated and not the data downloaded. This means that users who wish to reproduce this analysis will be able to do so provided they have $\sim 170$ GB of hard disk space available to store the results; a requirement that is met by most laptops nowadays. Second, when leveraging AWS to perform the analysis costs are minimized because there is no need to allocate additional Elastic Block Store (EBS) volumes to accommodate all $4.6$TB of darks. 
 
 \begin{figure}[H]
  \centering
  \includegraphics[width=\textwidth]{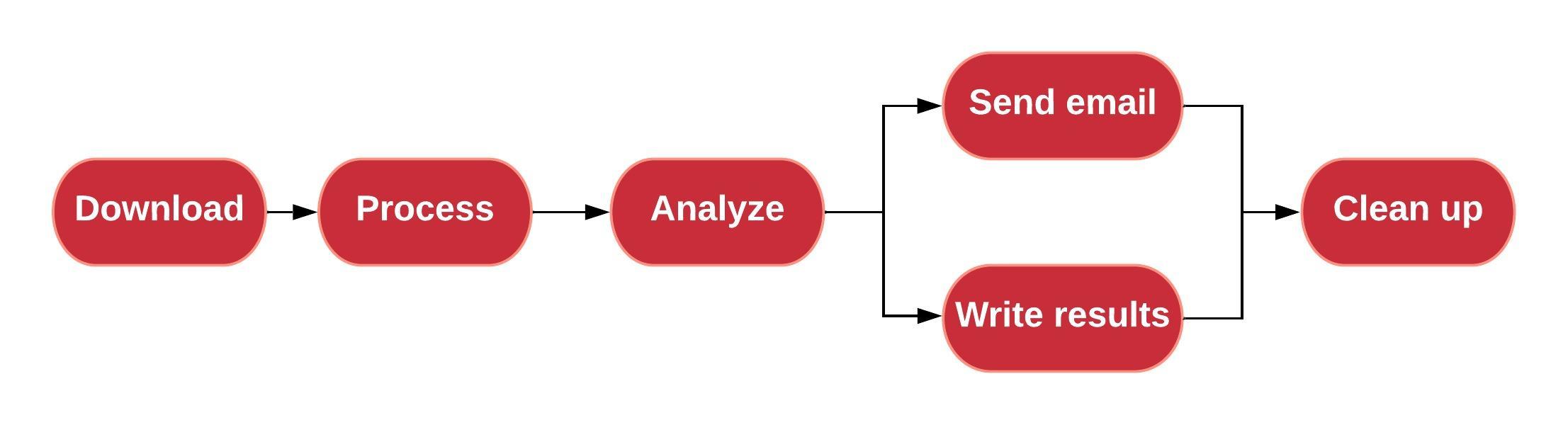} 
  \caption{Schematic overview of the processing and analysis pipeline.}
  \label{fig: pipeline_overview}
\end{figure} 
 
The pipeline consists of 5 distinct steps (Figure \ref{fig: pipeline_overview}) that are applied to consecutive one month intervals of darks. For each interval, we use \texttt{astroquery} \citep{astroquery} to programmatically query and download the bias-corrected dark frames along with  their engineering and telemetry files from the Mikulski Archive for Space Telescopes (MAST). Once downloaded, the images are processed through the cosmic ray rejection and identification steps. After processing, the identified cosmic rays are analyzed resulting in a catalog of parameters describing their morphology. After the analysis has completed, the results are written to file and an (optional) email is sent to the user defined email address with summary statistics for all the cosmic rays found in each image. Finally, all downloaded images and temporary files  are deleted to prepare for the next one-month chunk of darks.  

\subsection{Cosmic Ray Identification}\label{s:cr_algorithm}
\begin{figure}[H]
    \hspace{-2.5cm}
    \includegraphics[scale=0.28]{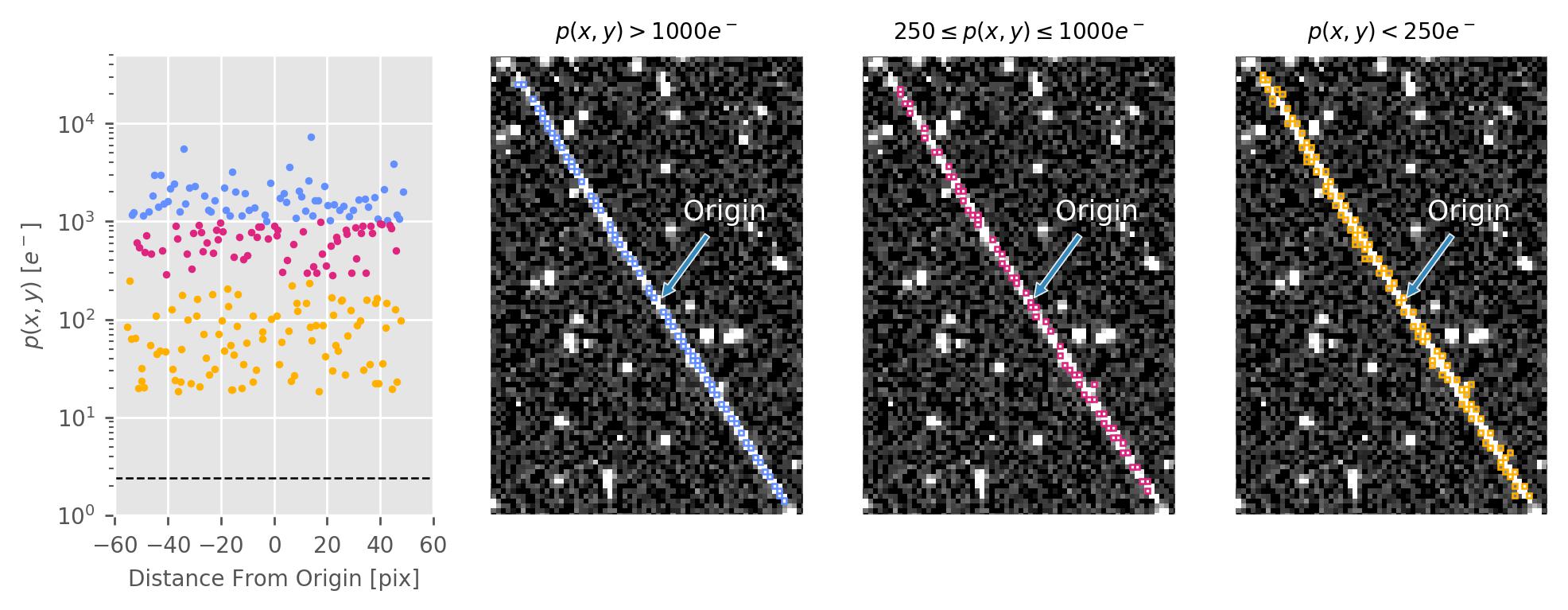}
    \caption{The scatter plot on the left shows the pixel values as a function of their distance to the origin marked in the image cutouts. We color code the points into three distinct groups; pixel values below 250 $e^-$ are yellow, pixel values between 250 $e^-$ and 1000 $e^-$ are magenta, and pixel values greater than 1000 $e^-$ are blue.  In the three image-cutouts, we show the location of pixels belonging to the three groups. The pixel values associated with this single cosmic ray span two orders of magnitude from 10's of electrons to 1000's of electrons and there is no obvious "profile" for how the electrons are distributed. By comparing the three bins of pixel values and their locations along the cosmic ray track, it is easy to see why any deblending algorithm will incorrectly segment this single object into numerous sources. }
    \label{fig:grazing_cr}
\end{figure}

Unlike other external sources (e.g. stars, galaxies), cosmic rays are unaffected by the optics of the telescope. Hence, there is no prescription for determining apriori how the energy deposited by a cosmic ray is distributed amongst the pixels that it affects. For unresolved sources, one can use knowledge of the PSF to determine how much flux, from two (or more) blended sources, is present in the pixels they share in common. For charged particles (e.g. cosmic rays), the energy deposited in a given pixel is probabilistic and so the energy received by adjacent pixels from the same cosmic ray fluctuates. In Figure \ref{fig:grazing_cr}, we highlight an example of this by showing an elongated cosmic ray observed in a STIS dark, o3sl01pcq\_flt.fits, taken in 1997.

For very elongated cosmic rays, these pixel-to-pixel fluctuations can be large enough such that typical deblending software (e.g. \texttt{Source Extractor}) will mistakenly segment a single elongated cosmic ray into numerous smaller ones.  In the same vein, spatially coincident cosmic rays lack the contrast required to reliably identify a local minimum in their combined source profiles making it extremely difficult to separate the two objects. For these reasons, we do not attempt to deblend spatially coincident cosmic rays as this does more harm than good by introducing a non-negligible number of false positives from oversegmentation which artificially raises the observed cosmic-ray particle flux.

In the following subsections we elaborate on the two techniques used to identify and label cosmic rays. We describe the process for instruments that are currently operationally active on HST, hereafter referred to as the active instruments, and those that are no longer operationally active, hereafter referred to as the retired instruments. The active instruments have robust calibration software that is available for use by the astronomical community through Python which greatly simplifies the process of identifying cosmic rays. The retired instruments lack similar calibration software and so a different method is utilized.

\subsubsection{Active Instruments}\label{s:active_instruments}

For the active instruments, ACS, STIS, and WFC3, we use their python packages, \texttt{acstools}, \texttt{stistools}, and \texttt{wfc3tools}, to run their cosmic ray rejection routines \texttt{ACSREJ}, \texttt{OCCREJECT}, and \texttt{WF3REJ}, respectively. As part of the cosmic ray rejection step, the Data Quality (\texttt{DQ}) extension of each input file is updated to indicate which pixels were affected by cosmic rays. Each routine implements a noise-based rejection model that looks for statistically significant outliers in observations made in sequence, at the same exact pointing. Because the images are taken at slightly different times and the same exact pointing, actual (non-transient) sources (stars and galaxies) will have the same detector position in each pointing, but cosmic rays and other transient artifacts will not. This allows for an easy identification of cosmic rays. Here we provide a review of the underlying algorithm utilized by the active instruments to identify pixels affected by cosmic rays.

The algorithm works by first performing a global background subtraction using the mode for each image. Next, the initial guess for the cosmic ray cleaned image is computed by taking the minimum or median across the stack of background subtracted images. For each image in the stack, a pixel-wise comparison with the initial guess is performed to determine which pixels have been affected by cosmic rays. If $p_n(x, y)$ is the value of the pixel at position $(x, y)$ in the $n^{th}$ image, $T_n$ is the exposure time of the $n^{th}$ image, $sky_n$ is the sky background of the $n^{th}$ image, and $p(x, y)$ is the value of the same pixel in the comparison image, then the variance with respect to the comparison image is computed as:
\begin{equation}
 \Delta_{n}(x,y) = \frac{[p_n(x,y) - (sky_n + p(x,y))]^2}{T_n^2}
 \label{eq:delta}
\end{equation}

This value is then compared to the expected variance, $\tau_n(x,y)$, for the given pixel,
\begin{equation}
\tau_n(x,y) = \left(\frac{\sigma^2}{T_n^2}\right) [\sigma_{RN}^2 + \sigma^2_{p}(x,y) + \lambda(p(x,y) - sky_n)^2],
\label{eq:tau}
\end{equation}
where $\sigma$ is a number representing the required level of significance (e.g 3 or 5), $\sigma_{RN}$ is the read noise of the amplifier used to read out the pixel at $(x, y)$, $\sigma_{p}(x,y)$ is the Poisson noise in comparison image, and the last term, $\lambda(p(x,y) - sky_n)^2$, is used to accommodate the undersampled PSF of the HST imagers. 

If $\Delta_{n}(x,y) > \tau_n(x,y)$, then the pixel at $(x,y)$ is marked with a special bit flag in the data quality (\texttt{DQ}) extension to indicate that it has been affected by a cosmic ray. If a pixel is identified as cosmic ray contaminated, the rejection criteria is applied to the neighboring pixels with a stricter $\sigma$ value. Finally, if multiple values are supplied for the $\sigma$ parameter, the algorithm will be applied in an iterative manner.
\begin{figure}[H]
    \centering
    \includegraphics[width=\textwidth]{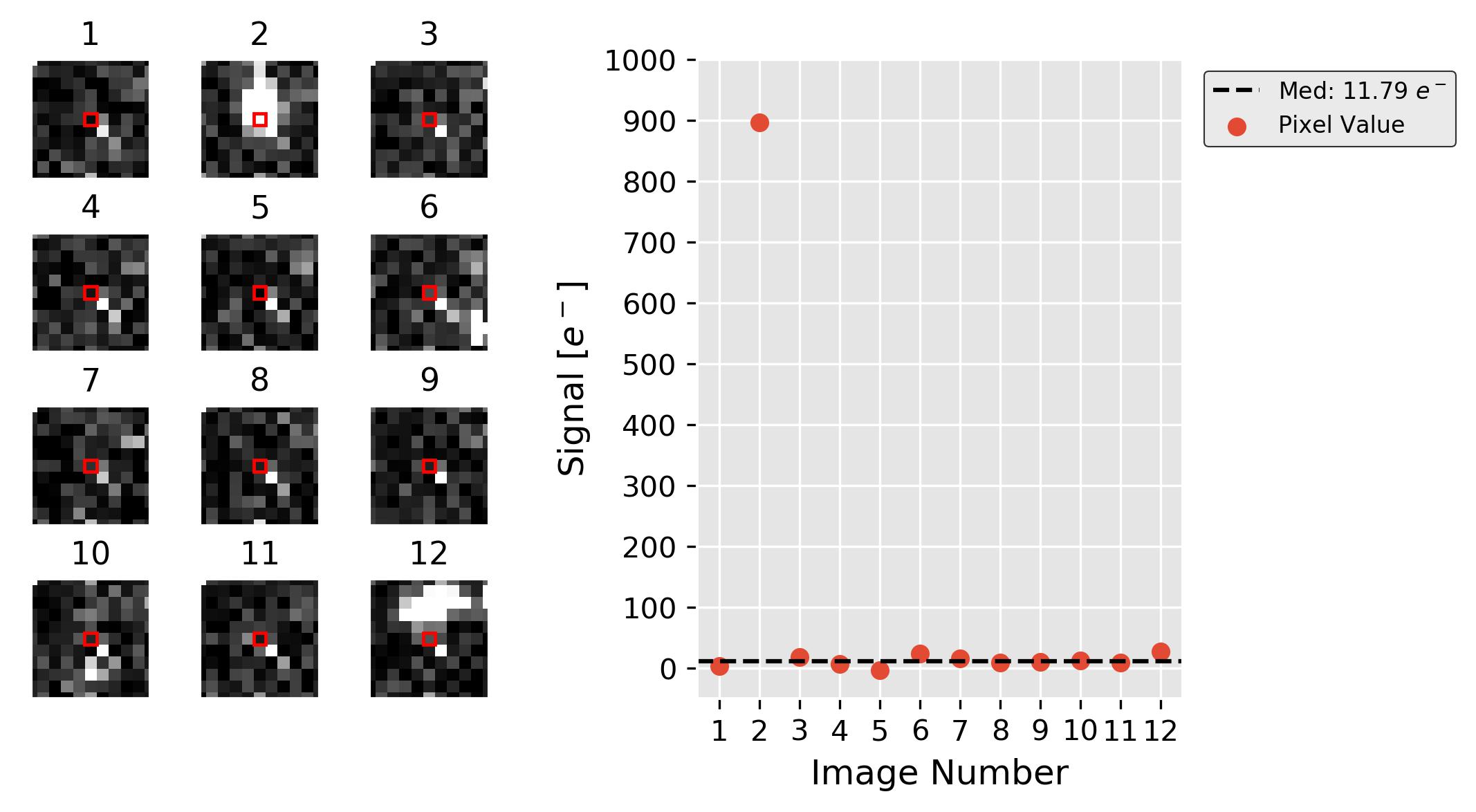}
    \caption{A visual representation of the cosmic ray rejection algorithm. Left: 10 pixel by 10 pixel cutouts from 12 different ACS/WFC images centered on the pixel marked by a red square. Right: The value of each pixel marked by the red square in the 12 different cutouts. The dashed vertical line is the median of all 12 values and represents the initial guess for the true value of the pixel marked by the red square. The only image where the pixel was marked as a cosmic ray by the ACS/WFC cosmic ray rejection routine, \texttt{ACSREJ}, is image number 2. }
    \label{fig:my_label}
\end{figure}

After running the cosmic ray rejection routine, 
the \texttt{DQ} extension of each input file has been updated with a special BIT flag, 8192, to indicate which pixels are 
affected by cosmic rays. Key to our work is utilizing this information to construct a binary image suitable for 
connected-component labeling analysis. We perform a bitwise-AND comparison between the \texttt{DQ} extension and 8192 to generate a binary image where any cosmic ray affected pixel is marked by 8192, everything else is marked with 0. We use the 8-connectivity matrix (Eq. \ref{eq: connectivity_matrix}) to identify all groups of connected pixels affected by the same cosmic ray. Any object identified that affects 1 or 2 pixels is rejected. This allows for a robust rejection of any unstable, hot pixels \citep{borncamp2017} identified during the cosmic ray rejection step 
which have large fluctuations in dark current that can be mistakenly classified as a cosmic ray by noised-based cosmic ray rejection algorithms like \texttt{ACSREJ}.

\begin{equation}
\begin{matrix}
 \begin{bmatrix} 
      1 & 1 & 1 \\
      1 & p & 1 \\
      1 & 1 & 1 \\
   \end{bmatrix}\\
  \end{matrix}
   \label{eq: connectivity_matrix}
\end{equation}

\begin{figure}[H]
    \centering
    \includegraphics[scale=0.15]{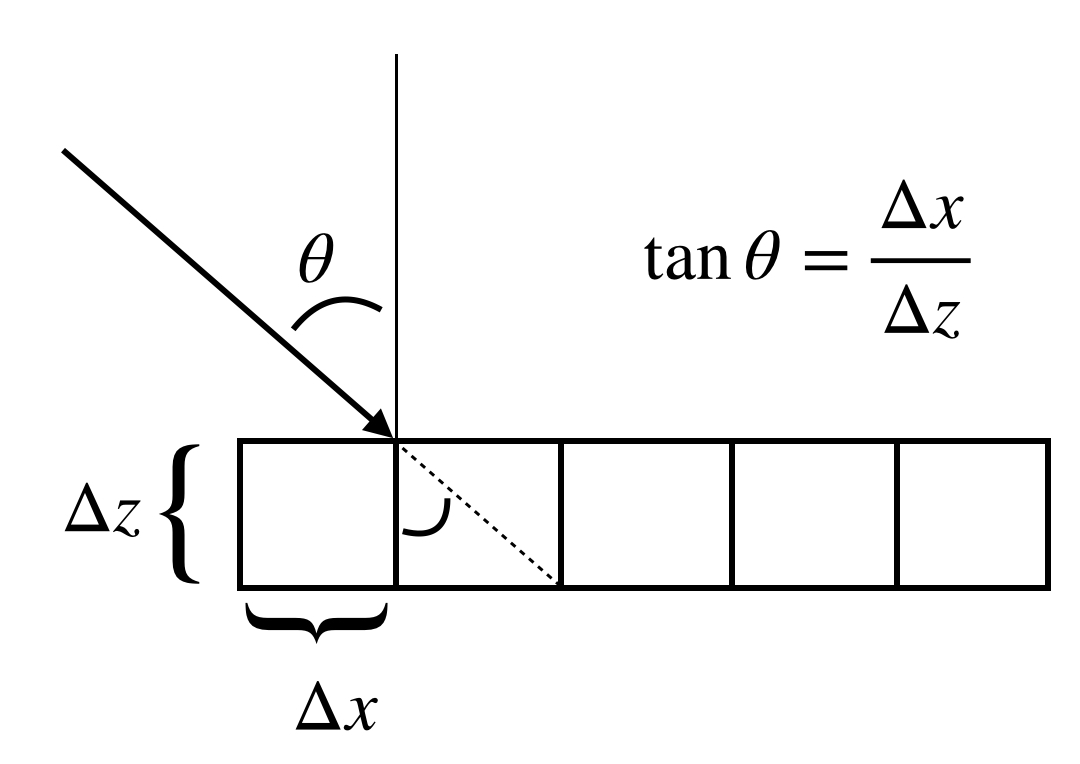}
    \caption{Cartoon depiction of the interaction between a cosmic ray and the pixel grid of a CCD. The trajectory of the cosmic ray is denoted by the arrow, $\Delta$$x$ is the pixel width, $\Delta$$z$ is the pixel thickness. The angle of incidence, $\theta$, is defined by the relationship in the figure.}
    \label{fig:cr_trajectory}
\end{figure}

\noindent
Filtering also removes potential single-pixel and two-pixel cosmic rays events. For a single-pixel event to occur on one of the HST CCD detectors, the charged particle (ignoring charge diffusion effects) must physically traverse only a single pixel. In Figure \ref{fig:cr_trajectory}, we show a cartoon depiction of the scenario where the path of the particle through the pixel is maximized. The thin, vertical line denotes the normal of the CCD's surface and it extends from the boundary between two pixels.

The particle trajectory denoted by the arrow will traverse a single pixel on an HST CCD if the angle incidence is in the interval defined by $[-\theta, 0)$ $\cup$ $(0, +\theta]$, where $\theta = \arctan{\Delta x/\Delta z}$. Note that for the case shown in Figure \ref{fig:cr_trajectory}, as $\theta$ approaches $0\degree$ the probability of electron-hole pairs generated in one pixel crossing pixel boundary via charge diffusion increases \citep{hopkinson1987} turning some single-pixel events into multi-pixel events.

The CCDs used by HST instruments are located off the main optical axis and pick-off mirrors are used to redirect light to each instrument housing. HST's pointing changes to observe astronomical sources, and during internal observations, e.g. dark frames, the telescope pointing is unconstrained. Hence the orientation of the CCDs with respect to the average direction of the particle flux is constantly changing. Further, cosmic rays originate outside the magnetosphere and so their trajectories are significantly altered by the geomagnetic field as they are confined move along field lines. These factors combine to significantly reduce the probability of  single- or two-pixel cosmic ray events in images taken with the HST CCDs. 

\begin{figure}[H]
    \centering
    \includegraphics[width=0.8\textwidth]{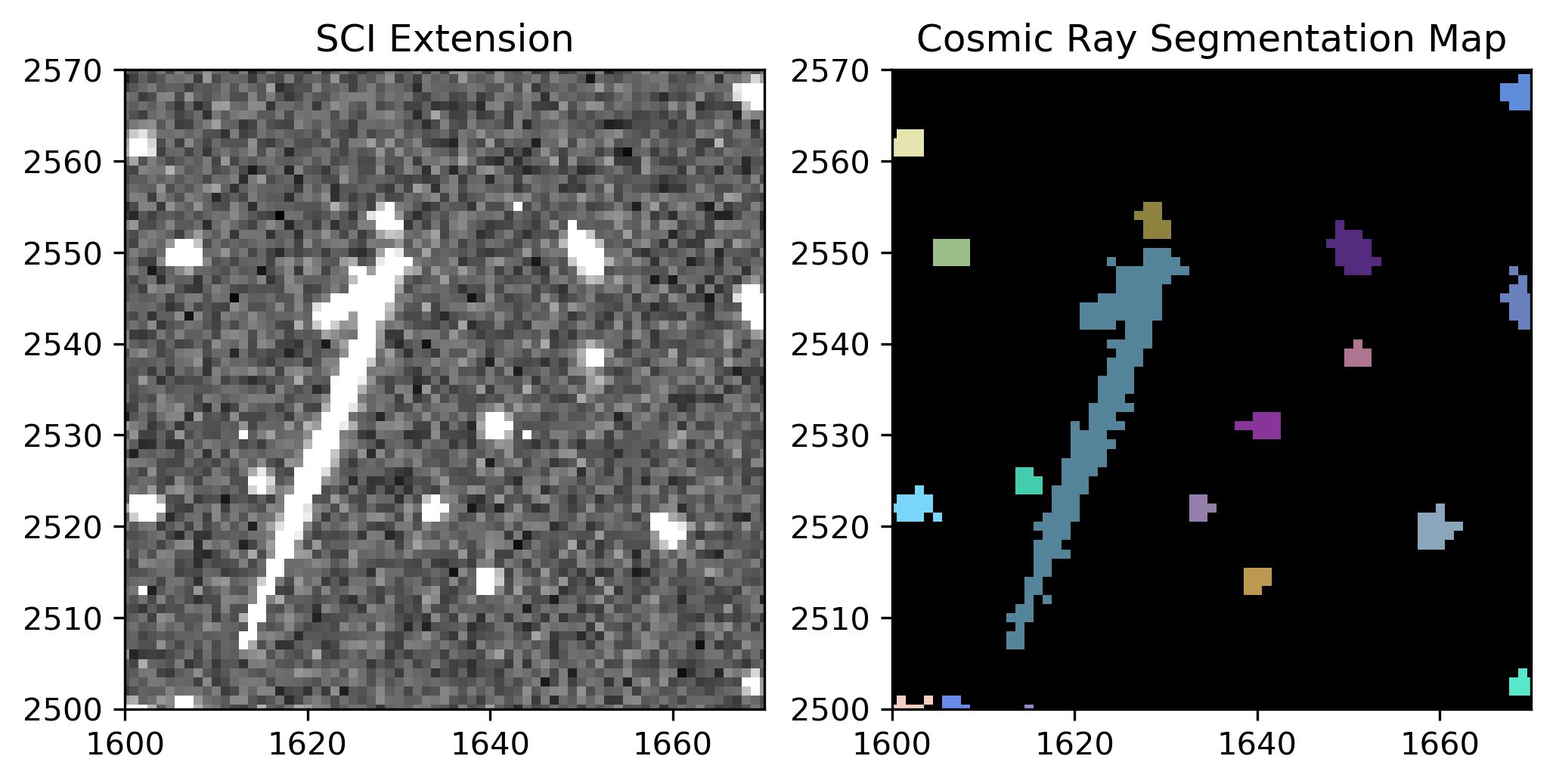}
    \caption{The \texttt{SCI} extension of an arbitrary STIS/CCD dark frame and the cosmic ray segmentation map produced by analyzing the \texttt{DQ} array associated with the \texttt{SCI} extension. As discussed in the beginning of this section, we do not attempt to deblend overlapping cosmic rays.}
    \label{fig: method2_label}
    \vspace{-0.2cm}
\end{figure}

In Figure \ref{fig: method2_label}, we show a cutout of the \texttt{SCI} extension and the corresponding segmentation map generated by labeling the \texttt{DQ} array for an arbitrary STIS/CCD dark. The colors indicate the distinct groups of pixels affected by individual cosmic rays. 

\vspace{2.1cm}
\subsubsection{Retired Instruments}\label{s:inactive_instruments}
\begin{figure}[H]
\vspace{-0.3cm}
    \centering
    \includegraphics[width=0.8\textwidth]{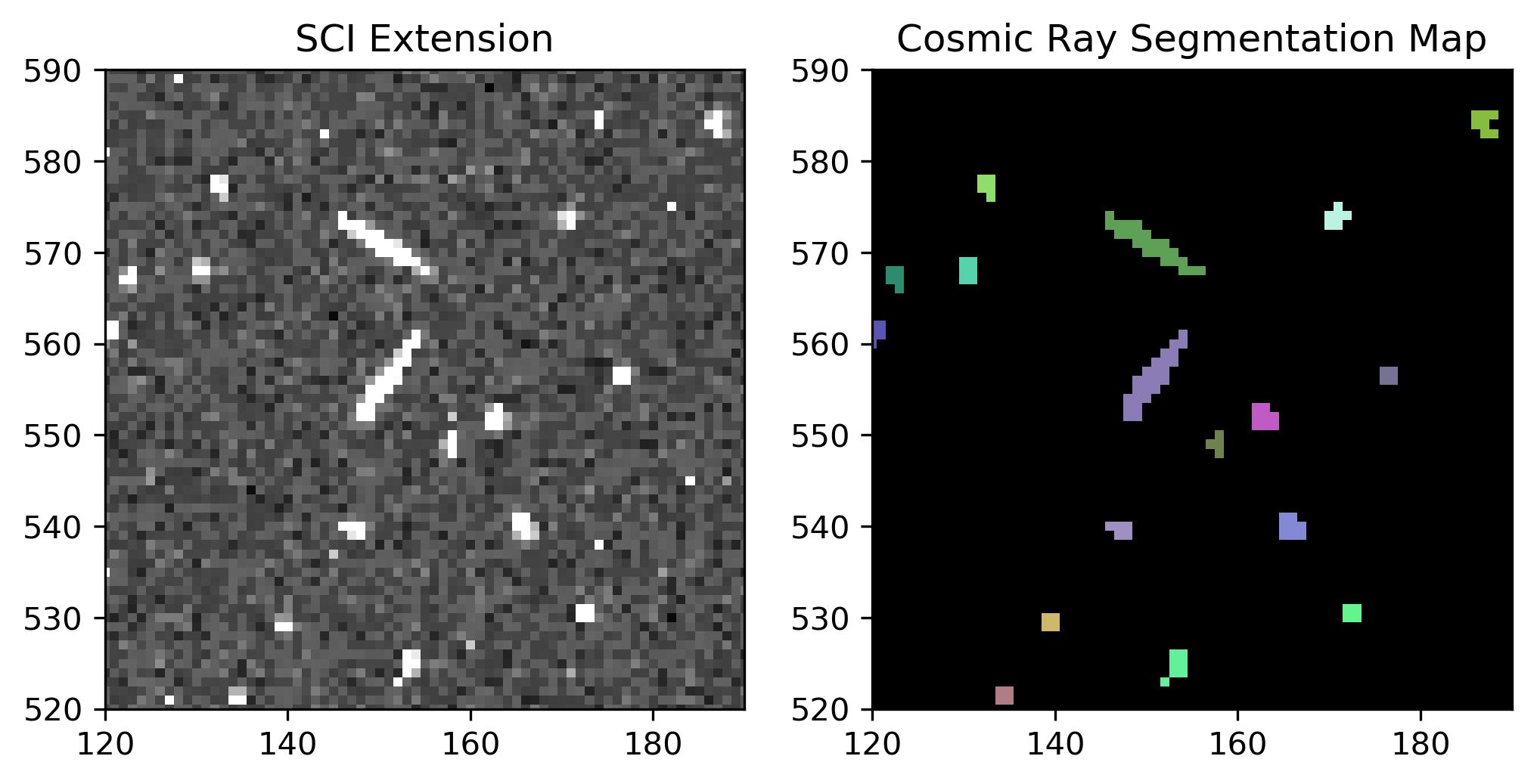}
    \caption{The \texttt{SCI} extension of a dark frame and the cosmic ray segmentation map produced by analyzing three sigma outliers in the \texttt{SCI} extenion. As expected, there are no hot pixels marked in the final segmentation map.}
    \label{fig: method1_label}
\end{figure}

Because the calibration software for the retired instruments, i.e. WFPC2, is deprecated, a different process is used. Each image is analyzed individually and a hybridization of binary thresholding and connected-component labeling, hereafter referred to as "threshold labeling", is used to identify cosmic rays. We perform three iterations of sigma clipping to compute the average pixel value, $\langle p \rangle$, and compute a robust measure of spread using the median absolute deviation (MAD). Using the sigma-clipped mean and the MAD, we create a binary image by marking all pixels according to the following condition:
\begin{equation}
     p(x,y) \geq \langle p \rangle + 5* \text{MAD} \begin{cases}
1 &\text{If the condition is True.}\\
0 &\text{If the condition is False.}
\end{cases}
\end{equation}
Next, we run the connected-component labeling analysis to identify groups of pixels with anomalously high values. We reject any object identified that affects 1 or 2 pixels to remove potential hot pixels and classify the rest as cosmic rays. We show an example of the resulting segmentation map generated by threshold labeling for WFPC2 in Figure \ref{fig: method1_label}.

\subsection{Comparing Threshold Labeling and LAcosmic}
To compare the results from threshold labeling and LAcosmic \citep{lacosmic} relative to the "ground truth" of the calibration pipeline's cosmic ray rejection algorithm we use dark frames obtained with ACS/WFC.

\begin{figure}[H]
    \centering
    \includegraphics[width=\textwidth]{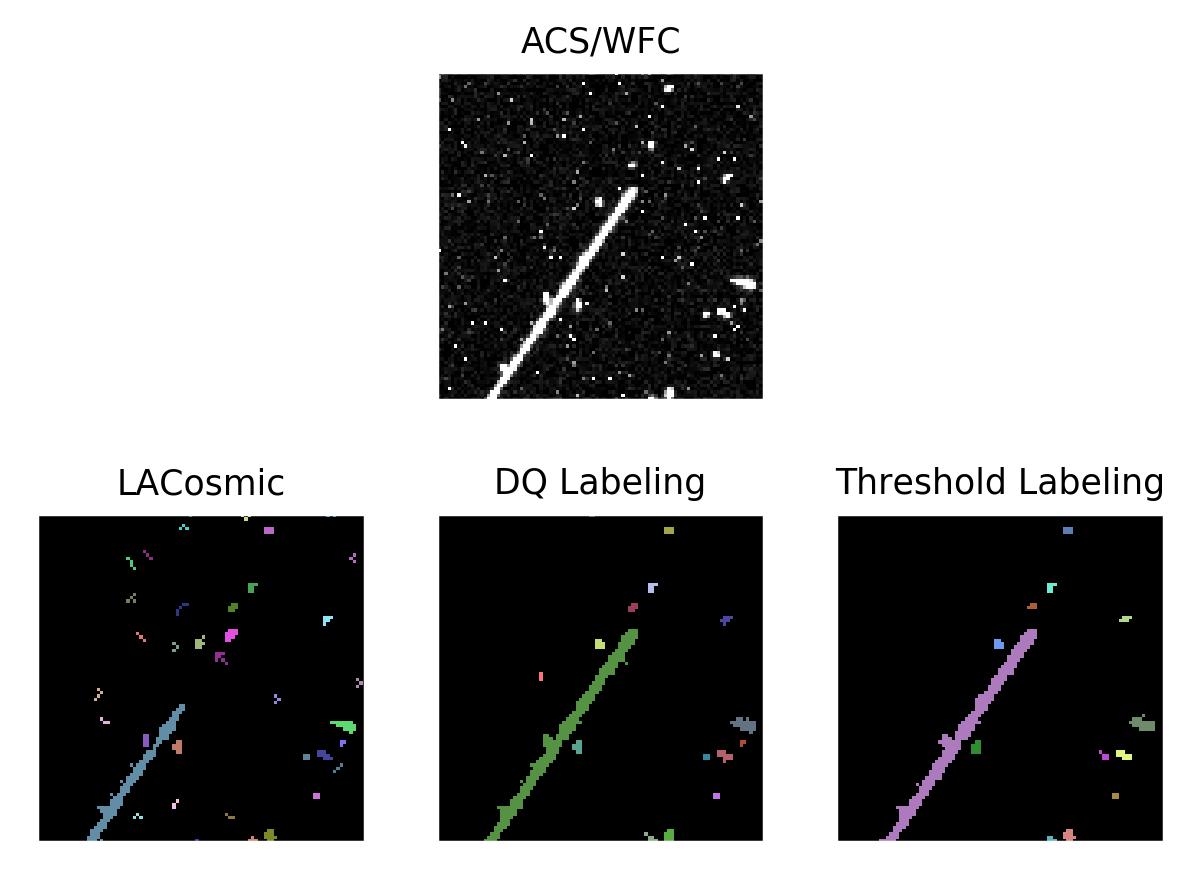}
    \caption{Visual comparison of the segmentation maps generated by the three methods. Top: A 100 pixel by 100 pixel subsection of the ACS/WFC image, j8jcnzv1q\_flt.fits. Bottom: The same region in the segmentation map generated by each algorithm.} 
    \label{fig:segmap_comp}
\end{figure}

We utilize the \texttt{astropy} \citep{astropy} implementation of LAcosmic, \texttt{astroscrappy}\footnote{\url{https://astroscrappy.readthedocs.io/en/latest/}}
\citep{astroscrappy}. Using data from the ACS/WFC, we run the cosmic ray rejection algorithm, \texttt{ACSREJ}, and use our method of labeling the \texttt{DQ} extension explained in section \ref{s:active_instruments} to define the "true" cosmic ray segmentation map. Next, we run LAcosmic with the default parameters to generate a cosmic ray mask and apply connected component labeling to create a second cosmic ray segmentation map. We generate a third cosmic ray segmentation map using the threshold labeling algorithm. In Figure \ref{fig:segmap_comp} we show a subsection of an ACS/WFC image and the corresponding segmentation maps generated by the three methods. 

We perform a pixel-wise comparison to determine the efficacy of each algorithm in identifying pixels affected by cosmic rays. We define $\mathbb{A}$ to be the set of all pixels identified by \texttt{ACSREJ} and $\mathbb{C}$ to be the set of all pixels identified by either LAcosmic or threshold labeling. We use the intersection, $\cap$, and set difference, $\setminus$, operators in conjunction with the sets defined above to compute the following two parameters,
\begin{equation}
  \alpha = \frac{\mathbb{C} \cap \mathbb{A}}{|\mathbb{C}|}
  \label{eq:alpha}
\end{equation}

\begin{equation}
  \beta = \frac{\mathbb{C} \setminus \mathbb{A}}{|\mathbb{C}|}
  \label{eq:beta}
\end{equation}

\begin{figure}[H]
    \centering
    \includegraphics[scale=0.3]{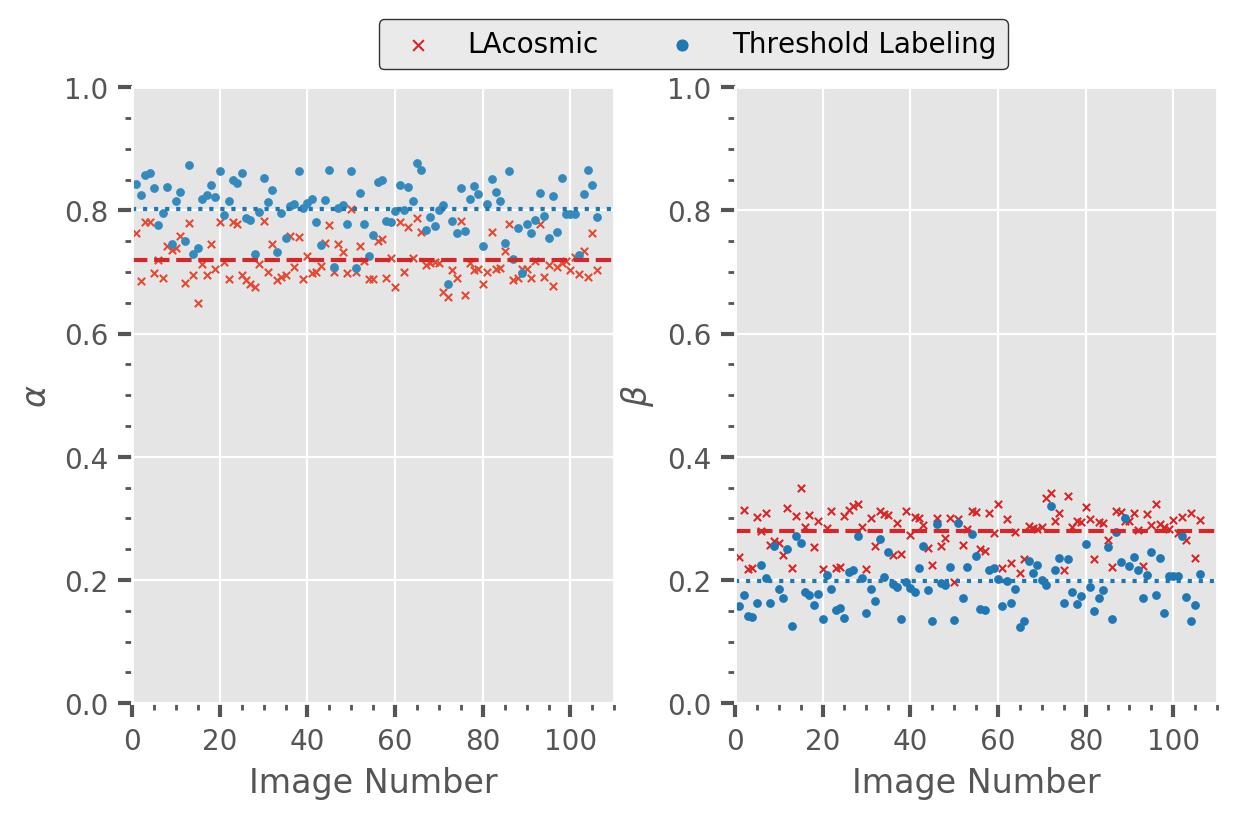}
    \caption{Left: The $\alpha$ value (Equation \ref{eq:alpha}) for 106 ACS/WFC images. Right: The $\beta$ value (Equation \ref{eq:beta}) for 106 ACS/WFC images.}
    \label{fig:frac_correct}
\end{figure}

The first parameter, $\alpha$, is the fraction of cosmic ray affected pixels that were correctly identified by each algorithm. The second parameter, $\beta$, is fraction of cosmic ray affected pixels that were incorrectly identified by each algorithm. In Figure \ref{fig:frac_correct}, we show $\alpha$ and $\beta$ for the 106 images analyzed. LAcosmic correctly identified an average of 72\% of the cosmic ray affected pixels and threshold labeling correctly identified an average of 80\%.    

We find that LAcosmic and threshold labeling identified an average of 15\% and 22\% more pixels, respectively, than \texttt{ACSREJ}. For LAcosmic, false positives accounted for nearly 30\% of all pixels identified, while for threshold labeling they accounted for 20\%. When analyzing darks with LAcosmic, individual hot pixels are routinely misclassified as cosmic rays because they have similarly "sharp" edges. These misclassifications result in a large number of false positives increasing the total number of detected cosmic rays by 66\% across the 106 dark frames.

The aim of our analysis is to study cosmic rays, so we adopt the threshold labeling algorithm for identifying cosmic rays in the WFPC2 darks. Compared to LACosmic, threshold labeling correctly identifies more true cosmic ray affected pixels with fewer false positives. However, because darks do not image external, astrophysical sources, our results should not be interpreted as the ability of LAcosmic to correctly distinguish between cosmic rays and 
external sources with broader profiles (e.g. stars), a regime where it excels.

\subsection{Analyzing Cosmic Rays}\label{s:analyzing_crs}
Once the cosmic ray label has been generated, we apply the label to the \texttt{SCI} extension to derive quantities of interest. For each cosmic ray we compute the following moments of the distribution of the deposited energy as defined in \citet{riess_2002},
\begin{itemize}
\item $I_0 = \sum_{i} p_i$
\item $I_x = \frac{1}{I_0} \sum_{i}p_i * x_i $
\item $I_y = \frac{1}{I_0} \sum_{i} p_i * y_i $
\item $I_{xx} = \frac{1}{I_0} \sum_{i}p_i(x_i - I_x)^2$
\item $I_{yy} = \frac{1}{I_0} \sum_{i}p_i(y_i - I_y)^2$
\item $I_{xy} = \frac{1}{I_0} \sum_{i}p_i(x_i - I_x)*(y_i - I_y)$
\end{itemize}

where $p_i$ is the pixel value of the $i^{th}$ pixel in the cosmic ray label and $x_i$, $y_i$ are the x and y coordinates of the $i^{th}$ pixel, respectively. The first parameter, $I_0$, is the total energy deposited by the cosmic ray in units of electrons. The second and third parameters combine to give the centroid of the cosmic ray, $(I_x, I_y)$. Using the second moments, $I_{xx}$ and $I_{yy}$, we compute the width or ``size" of the distribution of the deposited energy as, 
\begin{equation}
\text{size} = \sqrt{\frac{I_{xx} + I_{yy}}{2}}.
\label{eq:size}
\end{equation}
Lastly, we use the second moments to assess the symmetry of the distribution of the deposited energy,
\begin{equation}
\text{shape} = \sqrt{\frac{(I_{xx} - I_{yy})^2 + 4I^2_{xy}}{(I_{xx} + I_{yy})^2}}.
\label{eq:shape}
\end{equation}
For comparison, when applying Equation \ref{eq:size} and Equation \ref{eq:shape} to a 2D Gaussian with equal variance along both x and y (i.e. $\sigma_x=\sigma_y=\sigma$), Equation \ref{eq:size} yields $\sigma$ and Equation \ref{eq:shape} yields 0, indicating perfect symmetry \citep{hstcosmicrays_adass}. 

We compute the total number of pixels affected by each cosmic ray and record this as another metric for the "size" of the cosmic ray. We compute the cosmic-ray particle flux as the total number of individual cosmic rays identified divided by the total integration time and the size of the detector from Table \ref{tab: detectors}. The total integration time is defined as the exposure time plus half of the detector readout time which accounts for cosmic rays that strike the detector during readout. For each cosmic ray identified, we record the coordinates of all the affected pixels. 

Finally for each image analyzed, we use the engineering and telemetry files (*spt.fits) \citep{desjardins2019} to extract the following metadata:
\begin{itemize}
\item Altitude
\item Latitude
\item Longitude
\item Observation date
\item Observation start time
\item Observation end time
\item Telescope pointing (World Coordinate System (WCS) information)
\end{itemize}

\section{Results}\label{s:results}

\subsection{Cosmic Ray Morphology in HST Images}\label{s:morphology}
Cosmic ray morphological properties are of broad interest to the astronomical community because they can be used to discriminate astrophysical transients from cosmic ray events. In Table \ref{tab:cr_counts}, we report the total number of detected cosmic rays per instrument that we have analyzed thus far. 
\begin{table}[H]
    \centering
    \caption{The number of detected cosmic rays per instrument.}
    \begin{tabular}{cr}
    \toprule
      Instrument   & CR Count \\
      \midrule
      
      WFPC2     & 126,322,987 \\
      STIS/CCD  & 61,717,583 \\
      ACS/HRC   & 24,796,064\\
      ACS/WFC   & 558,517,641\\
      WFC3/UVIS & 526,545,187\\
    \midrule
    Total & 1,287,061,978\\
      \bottomrule
    \end{tabular}
    \label{tab:cr_counts}
\end{table}

\begin{figure}[H]
    \centering
    \includegraphics[scale=0.35]{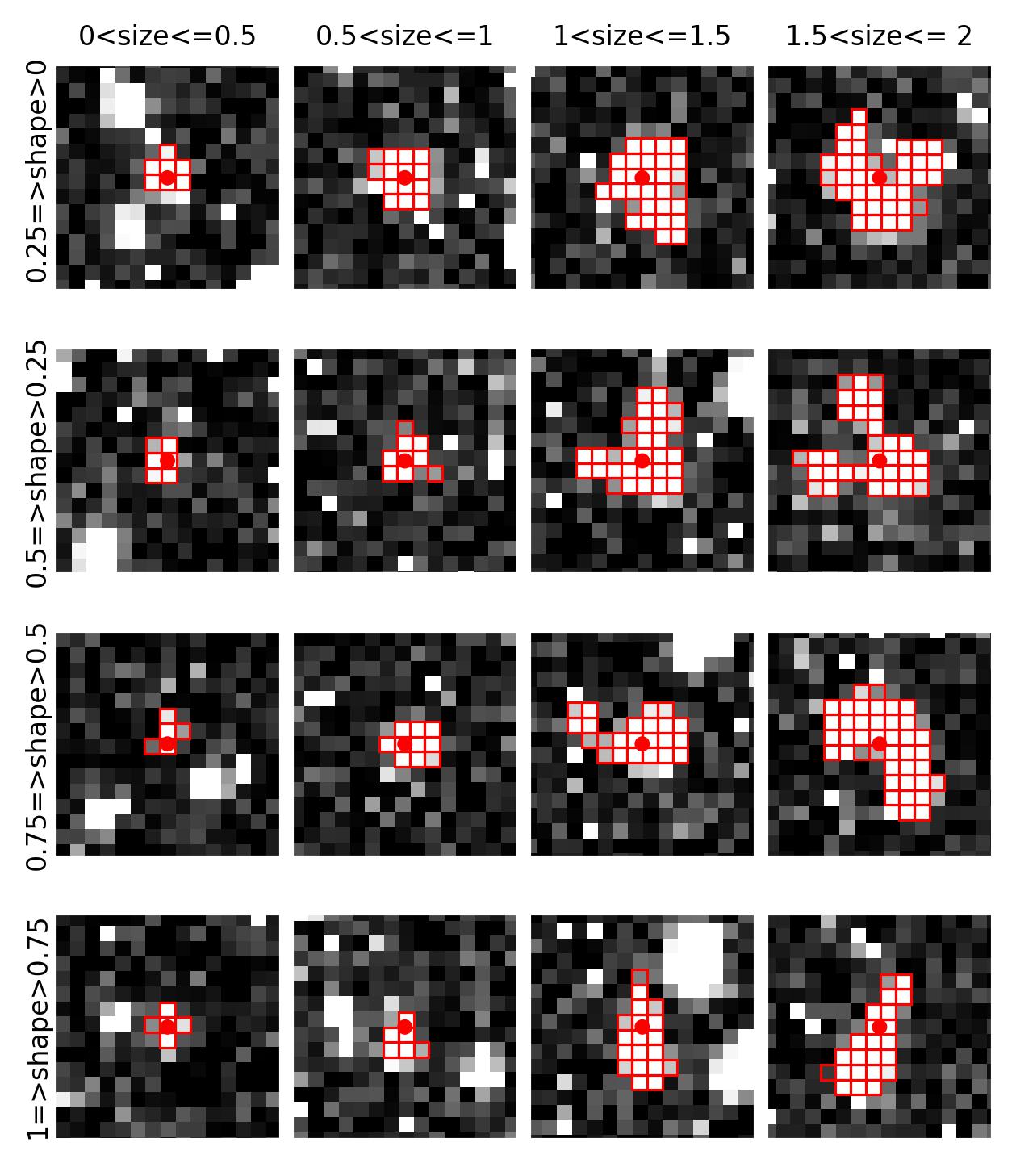}
    \caption{The morphology matrix for cosmic rays identified in a single ACS/WFC dark frame. In each subplot, the cosmic ray and the pixels it affected are marked by red squares. Each row corresponds to the shape bin denoted on the left, while each column corresponds to the size bin denoted on the top. The filled, red circle marks the computed centroid ($(I_x, I_y)$) of the cosmic ray.}
    \label{fig:crmorphology}
\end{figure}

In general, the morphology of cosmic rays in the two types of CCD detectors analyzed, i.e. thick, frontside-illuminated versus thin, backside illuminated, is highly consistent. In all the detectors they appear in a variety of shapes and sizes from elongated to point-like and are randomly distributed across the detector.
In Figure \ref{fig:crmorphology}, we show the morphology matrix for cosmic rays identified in a single ACS/WFC dark frame with an exposure time of 1000 seconds. The cosmic rays shown were randomly sampled from their corresponding distributions defined by the size and shape constraints for the given row and column. The top left corner corresponds to the smaller and more symmetric cosmic rays, while the bottom right corner corresponds to larger and more elongated cosmic rays. Where a given cosmic ray falls in the morphology matrix is almost entirely determined by its angle of incidence with respect to the normal of the plane of the CCD.

At nearly normal incidence, cosmic rays deposit energy in a symmetric manner leading to a small value for the shape parameter (Eq. \ref{eq:shape}) and a roundish appearance.  While at oblique incidence, energy is deposited asymmetrically resulting in large shape and size parameter values and elongated appearance. We postulate that objects in the lower left corner of the matrix are   the result of a very high energy cosmic ray interacting with atoms deep within the silicon substrate, far from gate that typically trap electrons generated by the photoelectric effect. When electrons are generated far from the gate, regardless of their origin, they are more likely to diffuse into adjacent pixels through a process known as charge diffusion \citep{hopkinson1987}.

A consequence of multiple angles of incidence, is that the distributions of the computed morphological parameters are highly asymmetric with a positive skew. 
Table \ref{tab:cr_statistics}
provides the summary statistics for each of the morphological parameters extracted. For each parameter, we report the $50^{th}$ percentile and the interval bounded by the $25^{th}$ and $75^{th}$ percentiles. The typical cosmic ray affects 7 pixels and deposits about 2700 electrons.

\begin{table}[H]
\centering
\caption{Summary statistics of the distribution of cosmic ray properties by detector.}
\begin{tabular}{crrrrr}
\toprule
{}	 &	WFPC2 &	STIS/CCD &	ACS/HRC &	ACS/WFC &	WFC3/UVIS\\
\midrule
Energy Deposited (e-) &\\

$25^{th} ~\%$	    &	1380.35	 &	1323.74	&	1469.49	&	1141.68	& 1323.78	\\
$50^{th}~\%$		&	2677.9	 &	2621.14	&	3024.74	&	1998.51	& 2239.04 	\\
$75^{th}~\%$	    &	16134.39 & 45958.78 & 115055.12 & 778211.98	& 108073.69	\\
Size Parameter$^{a}$ ($\sigma$) &\\

$25^{th}~\%$     &	0.438	&	0.426	&	0.421	&	0.481	&	0.477	\\
$50^{th}~\%$		&	0.536	&	0.512	&	0.493	&	0.591	&	0.602	\\
$75^{th}~\%$		&	0.928	&	2.004	&	1.156	&	1.674	&	6.135	\\
Size$^{b}$ (pixels)&\\

$25^{th}~\%$		&	4	&	5	&	5	&		6	& 6	 \\
$50^{th}~\%$		&	6	&	7	&	7	&		9	& 9	 \\
$75^{th}~\%$		&	10	&	39	&	13	&		20	& 46 \\
Shape Parameter$^{c}$ & \\

$25^{th}~\%$     &	0.423	& 0.356	& 0.326	&	0.339	& 0.365	\\
$50^{th}~\%$		&	0.646	& 0.582	& 0.523	&	0.538	& 0.575	\\
$75^{th}~\%$		&	0.915 	& 1.000	& 4.183	&	2.126 	& 1.542 \\	
\bottomrule
\end{tabular}
\begin{tablenotes}
\footnotesize
\item  $^{a}$ The size parameter $\sigma$ is derived using Equation \eqref{eq:size} and is equivalent to the gaussian standard deviation.
 \item $^b$ The number of connected pixels.  
\item $^c$ The shape parameter is calculated using Equation \eqref{eq:shape}
\end{tablenotes}
\label{tab:cr_statistics}
\end{table}%

\newpage
\subsection{Computing Particle Path Lengths}\label{s:path_lengths}

\begin{figure}[H]
    \centering
    \includegraphics[scale=1]{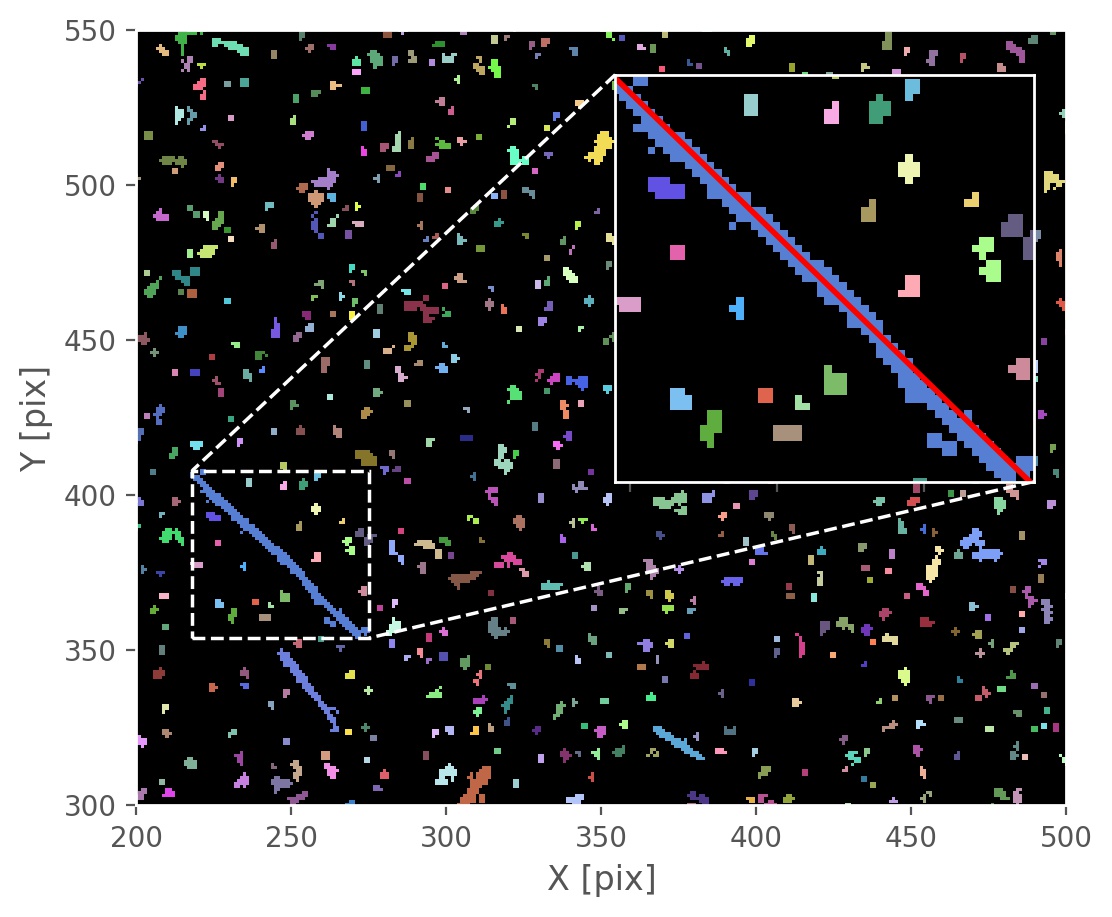}
    \caption{A visual representation of the projected path computed for a single, elongated cosmic ray in a reconstructed STIS/CCD cosmic ray segmentation map. The dashed, white box is the smallest box that fully encloses the elongated cosmic ray. The red line in the zoomed inset (the solid, white box) denotes the projected path length computed as diagonal of the box.}
    \label{fig:projected_path}
\end{figure}

The number of contiguous pixels affected by a single charged particle event represents the projected path length. For each image we reconstruct the cosmic ray segmentation map and identify the smallest box that encloses each cosmic ray. The diagonal of the box provides an estimate of the projected path length. Figure \ref{fig:projected_path} demonstrates this technique for one STIS/CCD dark frame. 

\begin{figure}[H]
    \centering
    \includegraphics[scale=0.22]{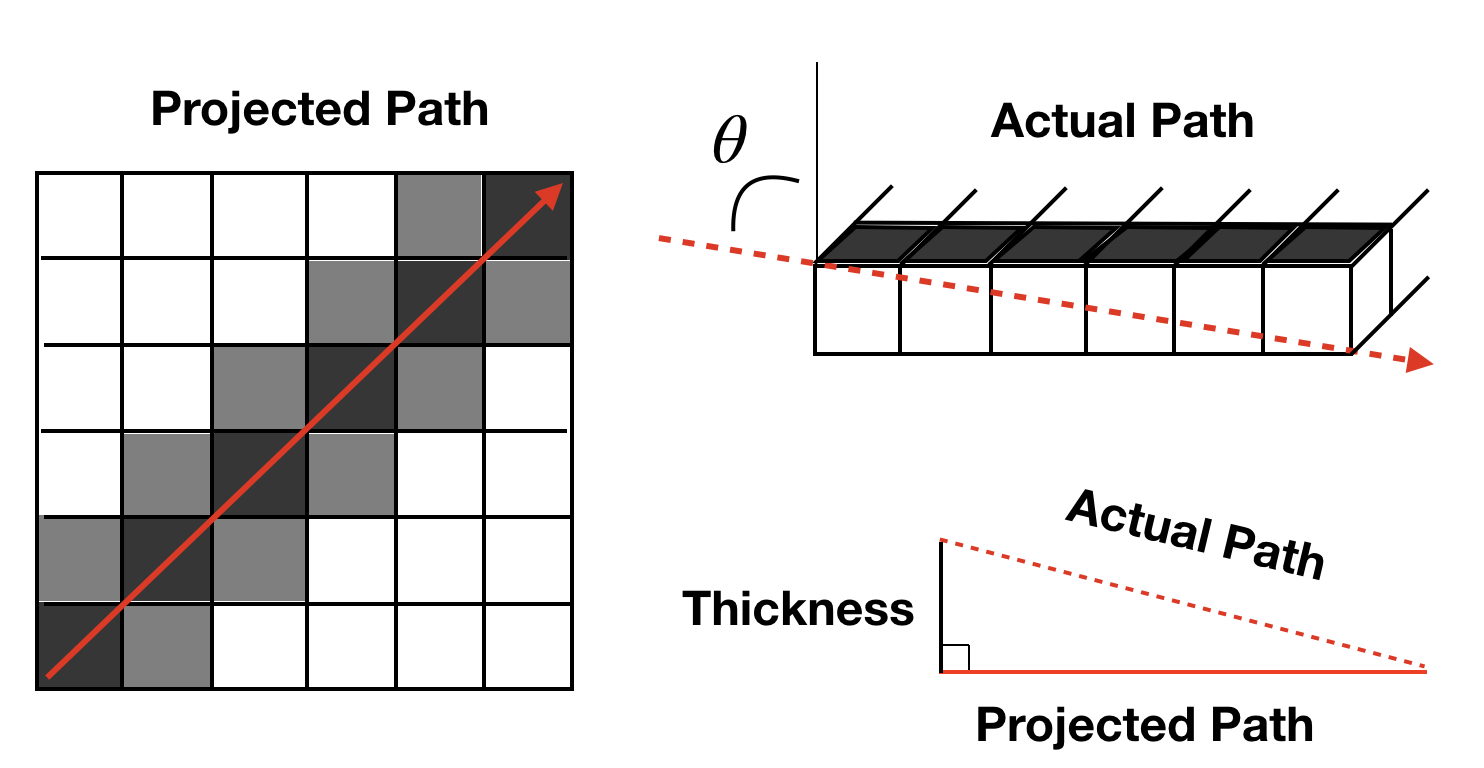}
    \caption{Schematic depiction of the method used to estimate the cosmic ray's actual path length through the CCD. Dark pixels have higher signal than white pixels, gray pixels are intermediate signal. }
    \label{fig:actual_path}
\end{figure}

Once the projected path length has been computed, we calculate the actual path through the detector as shown schematically in Figure \ref{fig:actual_path}. The projected path of the particle measured by the CCD pixels is shown on the left. The actual path 
\emph{through} the detector is shown on the right. We assume the pixel boundaries intersect the top and bottom surfaces of the CCD at 
90$\degree$ angles. Hence, the distance traversed through the detector is {$\text{Path Length}$ = $\sqrt{(\text{Projected Path})^2 + (\text{Thickness})^2}$}. 
We use the values listed in Table \ref{tab:ccd_props} to convert from units of pixels to units of micrometers. 
\vspace{-0.55cm}

\begin{table}[H]
\centering

\caption{Properties used to compute the path lengths through the CCD.}
\begin{tabular}{ccc}
\toprule
Instrument & Pixel Size  & Average Thickness [$\mu$m]  \\
\midrule
STIS/CCD   & 21 $\mu$m by 21 $\mu$m      & 14.04      \\
ACS/HRC    & 21 $\mu$m by 21 $\mu$m       & 14.26      \\
ACS/WFC    & 15 $\mu$m by 15 $\mu$m      & 14.85     \\
WFPC2      & 15 $\mu$m by 15 $\mu$m      & 10 \\
WFC3/UVIS  & 15 $\mu$m by 15 $\mu$m      & 15.75       \\
\bottomrule
\end{tabular} 
\label{tab:ccd_props}
\end{table}

The distribution of path lengths is most consistent with an isotropic particle flux modulated by strong shielding at large angles of incidence. Using the trigonometric relationships in Figure \ref{fig:actual_path}, the probability of a particle having a path length, $t$, is proportional to
\vskip-0.4cm
\begin{equation}
 A \sin(\theta)\cos^n(\theta)\frac{d\theta}{dt} = A/t^{(n+2)}
\end{equation}
where A is the normalization factor and \textit{n} represents the strength of shielding. The best-fit probability density function for the entire data set is $A/t^{(n+2)}$ with $n=2.33$, $A=3.83\mathrm{x}10^6$.

\begin{figure}[H]
    \centering
    \includegraphics[scale=0.85]{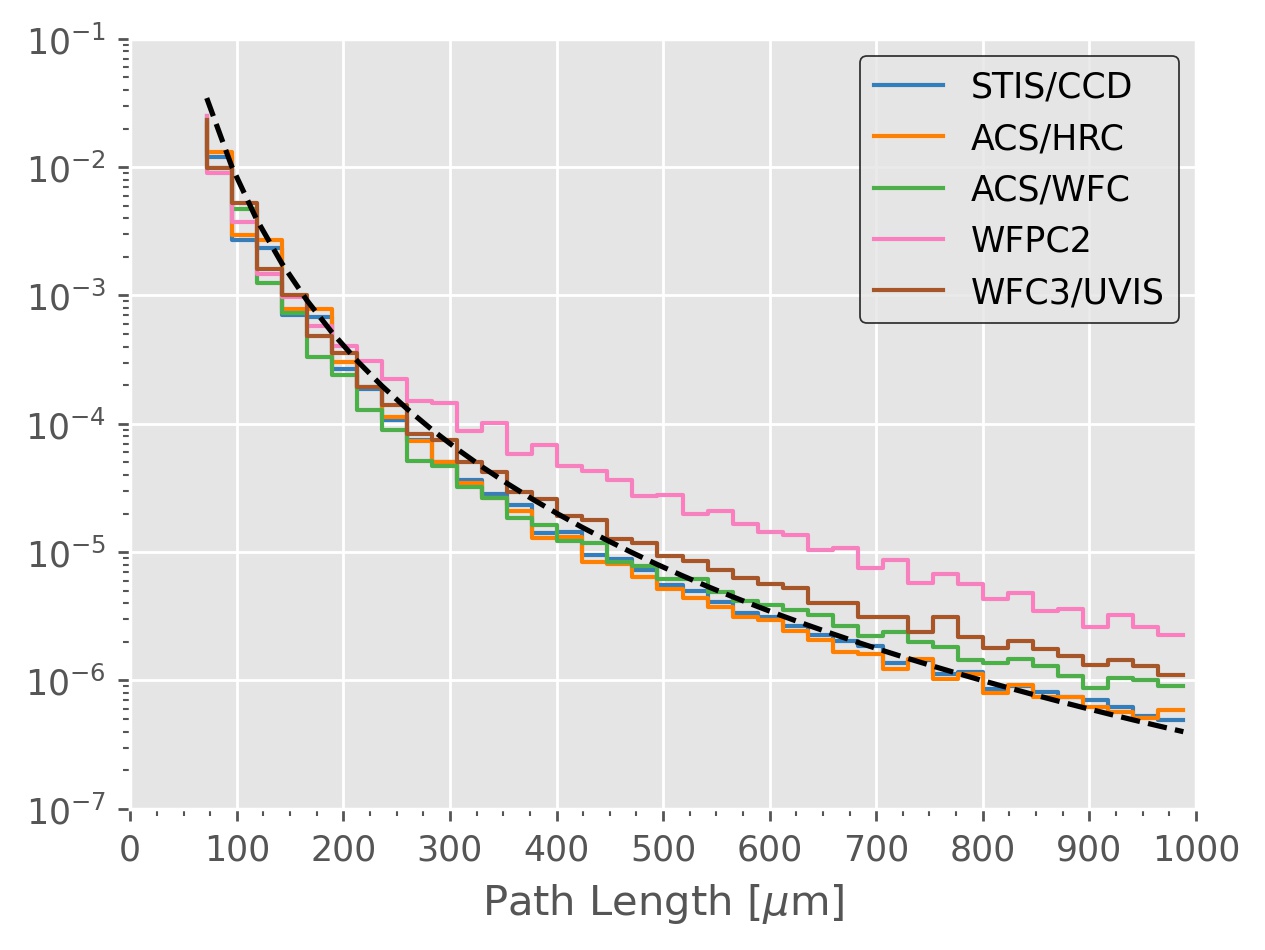}
    \caption{The distribution of path lengths for the 5 CCD imagers analyzed in this work. The dashed black line is the best-fit of the combined distribution of all 5 imagers.}
    \label{fig:path_length_fit}
\end{figure}

Figure \ref{fig:path_length_fit} shows the best-fit distribution, as well as the distribution of path lengths for each of the 5 CCD imagers. Although the distributions are similar, they are not identical and the distribution for WFPC2 notably deviates from that of the other imagers at 
around 200 $\mu$m, corresponding to a path length of about 13 pixels, possibly due to the location of its detectors in the observatory. 
\vspace{-0.25cm}

\newpage

\subsection{Cosmic-Ray Particle Flux}\label{s:cr_fluxes}

\begin{figure}[H]
    \centering
    \includegraphics[scale=0.8]{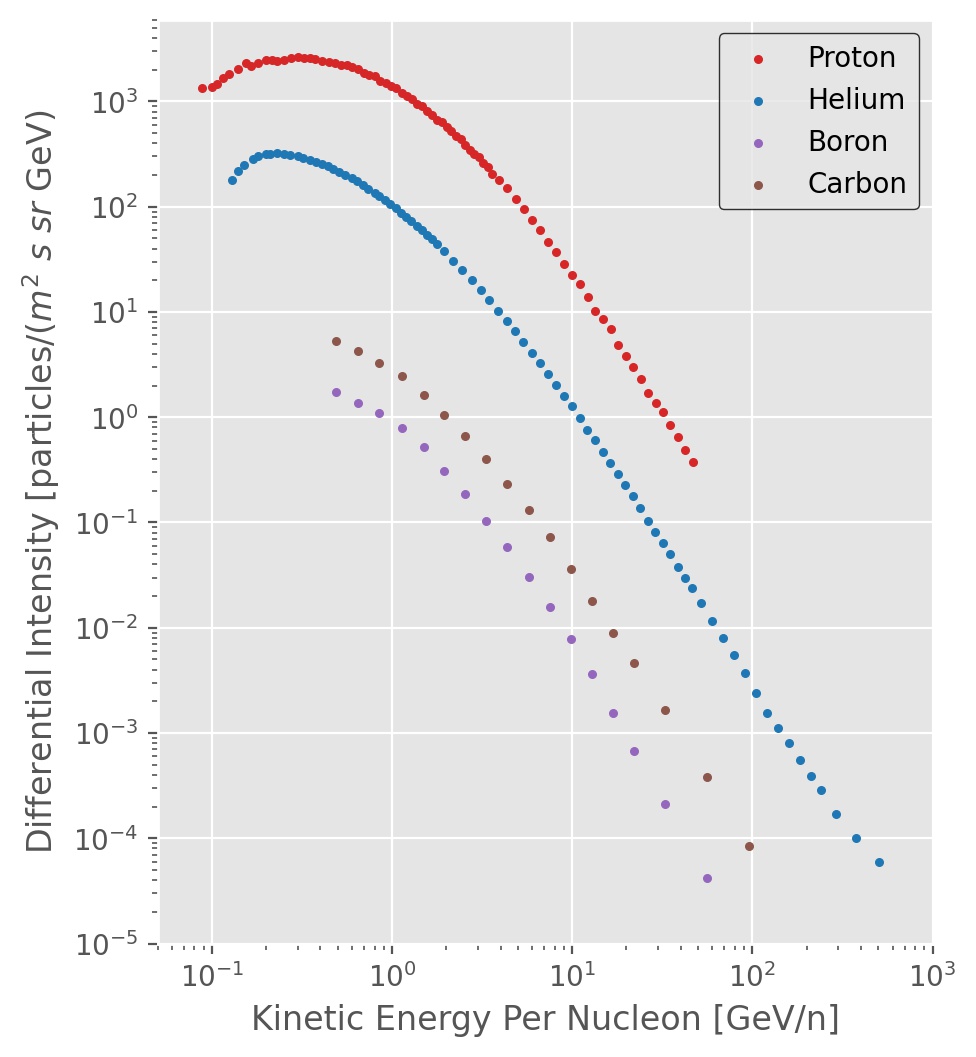}
    \caption{The differential intensity of cosmic rays observed by PAMELA. Data were obtained from the Cosmic Ray Database \citep{felice2017} hosted by the Space Science Data Center.}
    \label{fig:differential_intensity}
\end{figure}

The flux of cosmic rays at 1 AU in the solar system has been extensively studied by numerous ground- and space-based experiments. In Figure \ref{fig:differential_intensity}, we show the differential intensity of cosmic rays measured by the Payload for Antimatter Matter Exploration and Light-nuclei Astrophysics (PAMELA) experiment. The data were obtained from the Cosmic Ray Database \citep{felice2017} hosted by the Space Science Data Center. The flux of cosmic rays is primarily dominated by protons and helium at $\sim90\%$ and $\sim9\%$, respectively. The remaining 1\% is comprised of heavier nuclei and antimatter. While HST does not have a dedicated detector capable of distinguishing particle types, we may use the CCDs to estimate the bulk cosmic-ray particle flux in units of particles/($cm^2$ $s$) at HST's orbital altitude.

Within the observatory, each instrument is located in a distinct area of the telescope surrounded by differing amounts of shielding from space radiation. The amount of shielding around each instrument will determine the minimum energy a cosmic ray can have and still reach the CCD. The effects of this will be two fold. First, the low-energy regime of the power-law spectrum of cosmic-ray energies will be truncated as the low energy particles are completely absorbed by the shielding. Second, the inferred energy of cosmic rays that are detected will be less than their actual energy before encountering the shielding. This underestimation will result in a translation of the power-law spectrum shown in Figure \ref{fig:differential_intensity}, along the abscissa to lower values of the energy-per-nucleon and along the ordinate to smaller differential intensities. 


Additionally, the overall thickness varies from detector to detector and so a given cosmic ray will deposit more energy in a thicker detector. These differences affect the overall detection rates for cosmic rays. Thus to perform a direct comparison between instruments, an extensive analysis of the detector characteristics and the shielding for each instrument must be conducted, which is beyond the scope of this paper. 

\begin{figure}[H]
    \centering
    \includegraphics[scale=1.1]{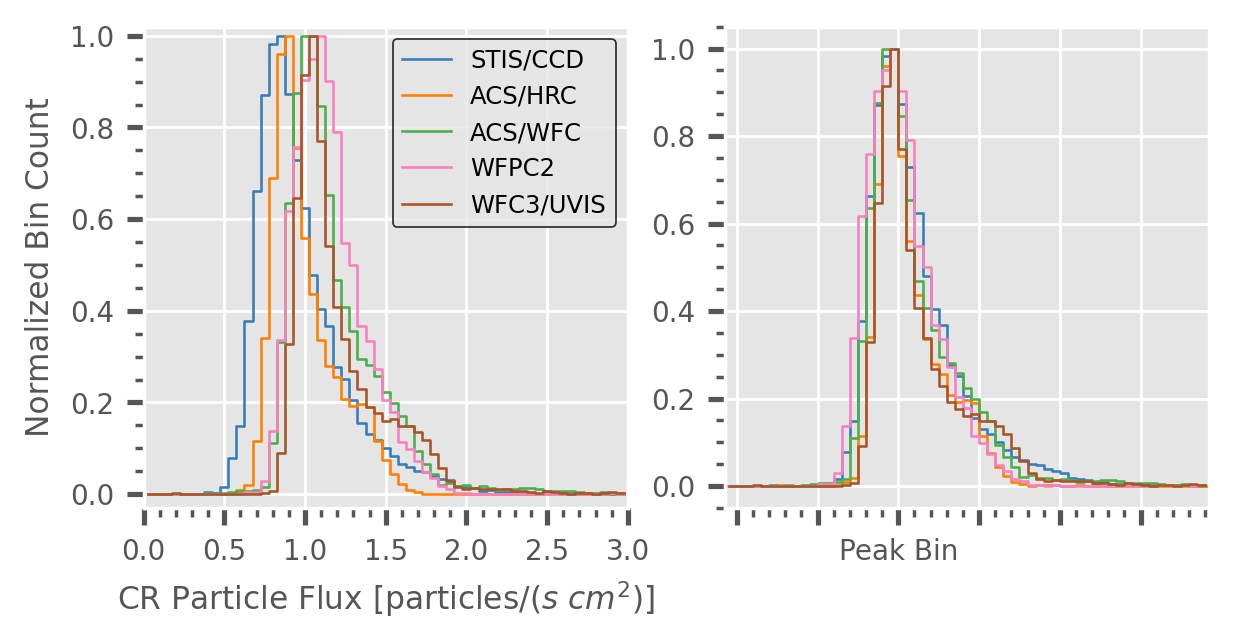}
    \caption{Left: The distribution of observed cosmic-ray particle fluxes for each instrument normalized to the peak bin. Right: The same distributions shifted so that their peak bins coincide to highlight the consistency of the overall distributions. }
    \label{fig:rate_hist}
\end{figure}

Nevertheless, in Figure \ref{fig:rate_hist} we show the distributions of the cosmic-ray particle flux for the imagers. We find that the distribution for each instrument has a positive skew and a well defined peak at $\sim$1 particle $s^{-1}$ $cm^{-2}$. In Table \ref{tab:cr_rate_statistics}, we report summary statistics computed for each distribution. 

\begin{table}[H]

    \centering
    \caption{Summary statistics describing the distribution of cosmic-ray particle fluxes for each instrument. The mean, error in the mean, and median are reported in units of \crpf {\deleted particles/$(s$ $cm^2)$}.}
    \begin{tabular}{ccccc}
	\toprule
	\centering
	{} &      Mean & Error in the Mean & Median & Num. Images\\
	\midrule
	ACS/HRC  & 1.013 & $2.55\mathrm{x}{10^-3}$ &  0.968  & 5297 \\
	ACS/WFC & 1.165 & $1.92\mathrm{x}{10^-3}$  & 1.123  & 12806  \\
	STIS/CCD & 0.956 & $1.21\mathrm{x}{10^-3}$  & 0.931  & 29599 \\
	WFC3/UVIS & 1.199 & $1.90\mathrm{x}{10^-3}$  & 1.146 & 12672 \\
	WFPC2 & 1.189 & $1.77\mathrm{x}{10^-3}$ & 1.167 & 13015  \\
	\bottomrule
	\end{tabular}
\label{tab:cr_rate_statistics}
 \end{table}

As a check on our calculated cosmic-ray particle fluxes, we compare our results to the flux estimated using the PAMELA data shown in Figure \ref{fig:differential_intensity}. PAMELA was a dedicated cosmic ray detector placed in low earth orbit, at an average altitude of about 500 km and with an orbital inclination of 70$\degree$. Note that while HST and PAMELA share a similar orbital altitude, the higher inclination of PAMELA implies a lower geomagnetic cutoff rigidity (see e.g., Figure 7 in \citet{smart2005}). Thus we should expect the PAMELA cosmic-ray particle flux to be slightly larger than that observed by HST. 

To determine the expected cosmic-ray particle flux, in units of $\sim$1 particle $s^{-1}$ $cm^{-2}$, we make some simplifying assumptions. First, we focus our analysis on the differential intensity of protons as they comprise the bulk majority of cosmic-ray particle types. Second, we assume their distribution is isotropic. With these assumptions, we estimate the cosmic-ray particle flux as,

\begin{align}
    F ={}& 2 \int^{E_{max}}_{E_{min}} \int^{2\pi}_{0} \int^{\pi/2}_{0} I(E)\cos{\theta}\sin{\theta} \,dE\,d\phi\,d\theta \\
    F ={}& 2\pi \int^{E_{max}}_{E_{min}} I(E) \,dE
\end{align}

We solve this integral numerically. First, we use cubic spline interpolation to sample the observed distribution at 0.1 GeV resolution. Next we perform the integration using the sampled datapoints over the range of kinetic energies in the PAMELA dataset, $E_{min}=0.1$ GeV to $E_{max}=46.6$ GeV.

We find that the flux of cosmic ray protons observed by PAMELA is 2.5 $\sim$1 particle $s^{-1}$ $cm^{-2}$, which is within a factor of 2.5 of the value derived in this work. If the minimum energy range is restricted to 1 GeV, a more realistic value for HST's low inclination orbit, the resulting integral yields 1.3 $\sim$1 particle $s^{-1}$ $cm^{-2}$. The average of the mean cosmic-ray particle fluxes reported in Table \ref{tab:cr_rate_statistics} is 1.11 $\sim$1 particle $s^{-1}$ $cm^{-2}$. These values are within 17\% of one another and demonstrate the capabilities of HST as a particle detector. 

\subsection{Probing Detector Thickness}\label{s:thickness}
When a visible light photon strikes the $Si$ detection layer of a CCD it is readily absorbed within a short distance due to the extremely high absorption coefficient of $Si$. However, the absorption coefficient is wavelength dependent and at longer wavelengths $Si$ becomes increasingly transparent (\cite{silicon_prop}). Since CCDs are comprised of layers of different $Si$ compounds, each with a different index of refraction, a fraction of the incident light will be transmitted and the remainder reflected at the boundary layer between two compounds. This reflected light can then interfere with any additional light entering to produce a fringe pattern. 

\begin{figure}[H]
\hspace{-0.85cm}
   \includegraphics[scale=0.27]{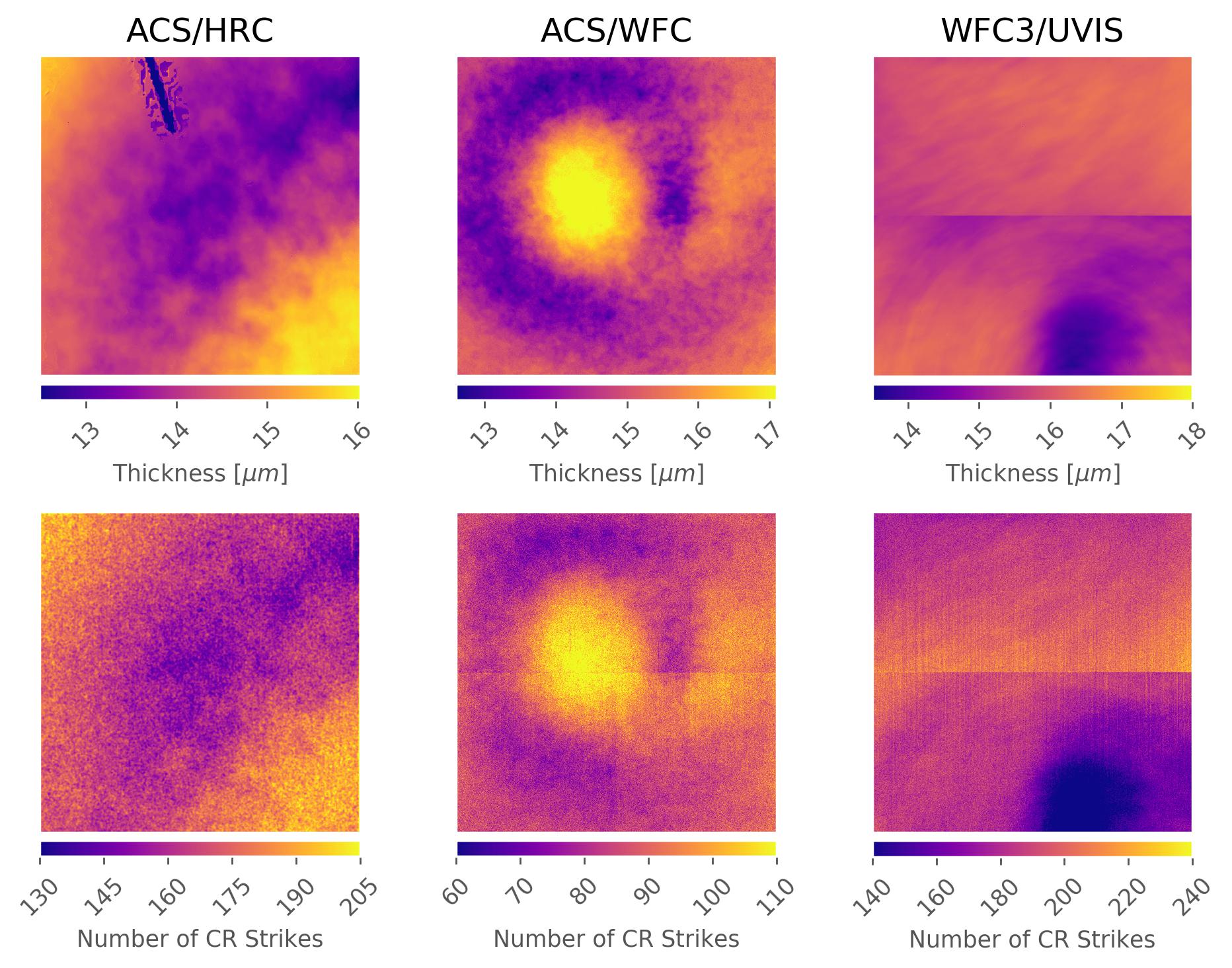} 
   \caption{Top row: The thickness maps produced by analyzing fringe patterns in the respective instruments. The ACS data was provided by J.R. Walsh from \citet{walsh_2003} and the WFC3 data was provided by M. Wong from \citet{wong_2010}. Bottom row: The corresponding cosmic ray heat maps. Note that by the nature of the cosmic ray heat maps, we were able to probe the region of the ACS/HRC detector obscured by the coronagraphic finger in the thickness map generated from the fringing analysis. }
   \label{fig: thickness_maps}
\end{figure}

\citet{malumuth_2003}, \citet{walsh_2003}, and \citet{wong_2010} modeled fringing in STIS, ACS, and WFC3 CCDs, respectively, to correct for the effect of fringing in a given observation. $Si$'s wavelength dependent absorption distance is well known, so by modeling the fringing, they derive the $Si$ detection layer thickness and produce a corresponding thickness map for each detector (Figure \ref{fig: thickness_maps}, top row). Using the record of pixels affected by cosmic rays, we generate a heat map of cosmic ray strikes for each CCD. This serves as a proxy for the thickness of the detector as thicker areas of the detector have more potential scattering targets. Comparing thickness maps produced from the fringing analyses to the cosmic ray heat maps, we show in Figure \ref{fig: thickness_maps} that we reproduce the overall detector structure. 

\subsection{Solar Modulation of the Cosmic Ray Flux}\label{s:solar_modulation}

The periodic nature of solar activity, first observed by counting sunspot numbers, has been extensively monitored since the 1800's and independently confirmed through a variety of other measures of solar activity (e.g. 10.7cm solar flux, total solar irradiance; see \citet{hathway_2015}). These observations led to the discovery of the 11-year sunspot and 22-year magnetic cycles. The 11-year cycle is defined by the time it takes for the total number of sunspots to progress from a minimum through a maximum, to the next minimum with each cycle begins at a the minimum \citep{russell2019}. The 22-year cycle, also known as Hale's Polarity Law, is the time for sunspot pairs to achieve the same magnetic polarity (\cite{hale_1919}) with respect to the rotational axis of the sun. In three consecutive sunspot cycles the first cycle will have sunspot pairs with a given polarity, the second will have sunspot pairs with the opposite polarity, and the last will have sunspot pairs with the same polarity as the first.

\begin{figure}[H]
    \hspace{-1.5cm}
    \includegraphics[scale=0.85]{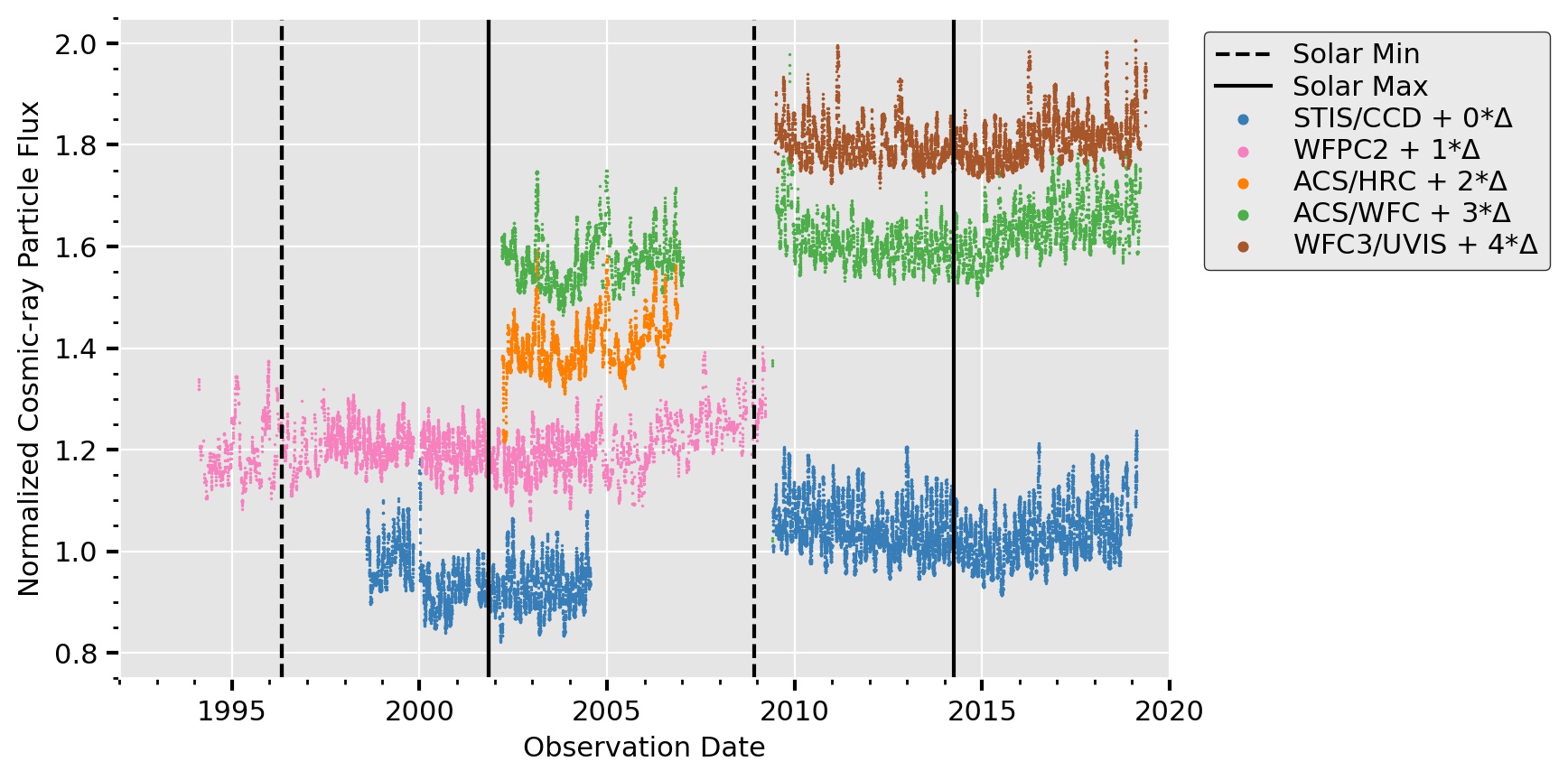}
   \caption{Top: The observed cosmic-ray particle flux for each instrument after being smoothed by a 1 month (30 day) rolling median and normalized by the median flux observed by the instrument. For clarity we use a constant offset, $\Delta=0.2$, to shift the normalized fluxes along the y-axis. The dashed vertical lines correspond to solar minima, whereas the solid lines correspond to solar maxima. }
   \label{fig:cr_rate}
\end{figure}

It is well known that the observed galactic cosmic ray (GCR) flux measured at the Earth is anti-correlated with solar activity \citep{potgieter_rev2013}. In Figure \ref{fig:cr_rate}, we show the median-normalized cosmic-ray particle fluxes measured by HST. We filter out observations in the SAA and smooth the time-series with a 30 day rolling median filter.

\vspace{-0.4cm}

Qualitatively, the variation of the observed cosmic-ray particle flux with the solar cycle is apparent.  About one year after each minimum (maximum), the cosmic-ray particle flux reaches its maximum (minimum).  The delay between the cosmic-ray particle flux and solar cycle extrema is expected as the response of the heliosphere to changes in solar activity is not instantaneous. The observed anti-correlation suggests that the majority of cosmic rays detected by HST are galactic in origin. 

\begin{figure}[H]
\hspace{-1.5cm}
    \includegraphics[width=1.2\textwidth]{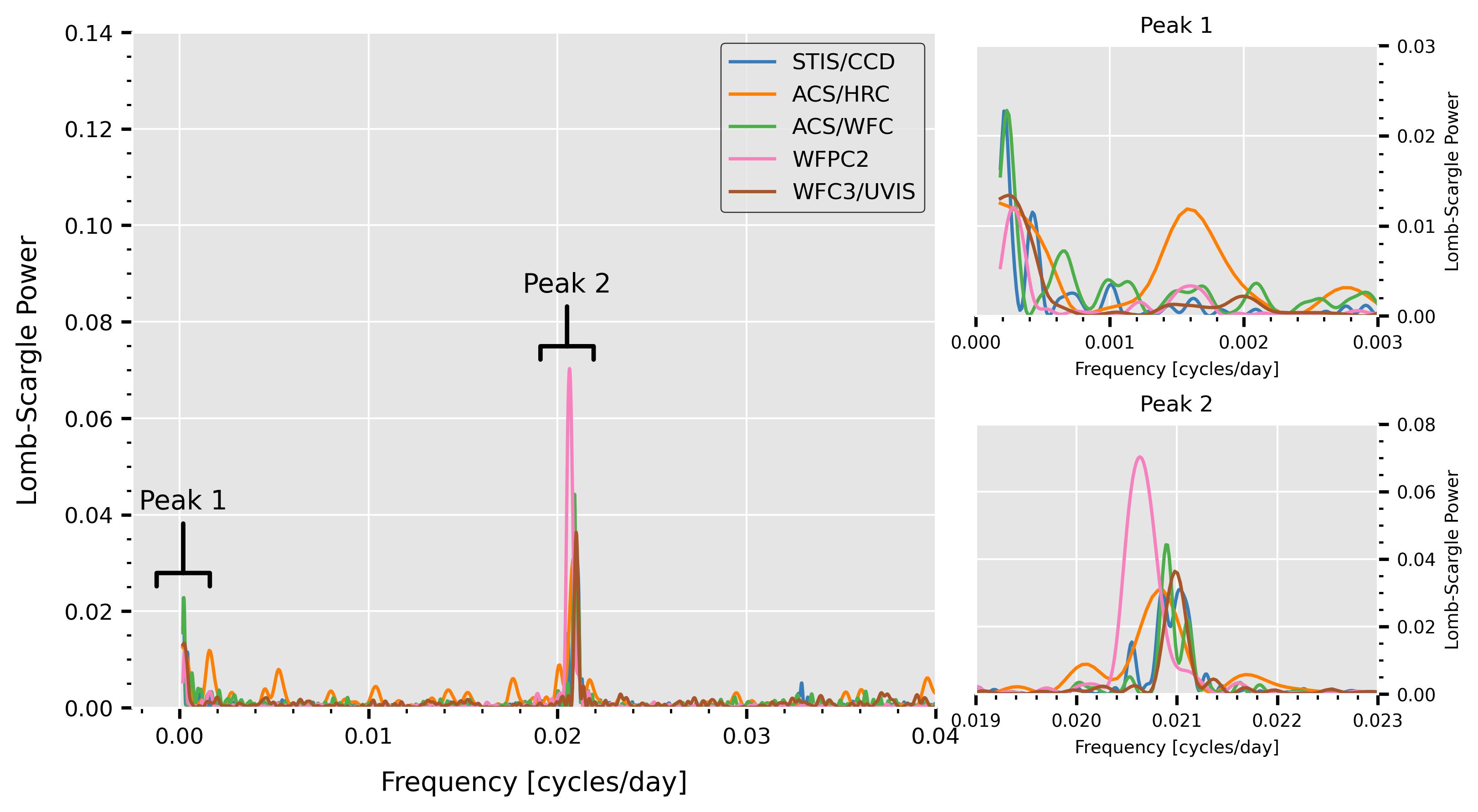}
        \caption{Left: The Lomb-Scargle periodogram for each of the 5 CCD imagers analyzed. Top Right: Same data, zoomed in to highlight the first peak. Bottom Right: Same data, zoomed in to highlight the second peak.}  
    \label{fig:lomb_scargle}
\end{figure}

Quantitatively, we perform a spectral analysis by computing the Lomb-Scargle periodogram of the observedcosmic-ray particle flux as a function of time (see Figure \ref{fig:lomb_scargle}). The results are similar for the five instruments, with the exception of the ACS/HRC which only operated for three years.  Its results are therefore unreliable (see Table \ref{tab: detectors}). The first peak occurs at $\sim 0.00024$ cycles/day corresponding to the 11-year solar cycle. The second peak at $\sim 0.021$ cycles/day corresponding to 48 days 
For each peak, the average false alarm probability is $\approx$ 0.

On average we find that the cosmic-ray particle flux varies by about 25\% between solar minimum and solar maximum across the detectors, consistent with values from Alpha Magnetic Spectrometer (AMS). \citet{corti2019} analyzed seven years of AMS\footnote{In 2011 AMS was installed on the International Space Station, which, orbits at 450 km above Earth's surface} proton flux data in three rigidity bins: 1.01 $-$ 1.16 GV, 4.88 $-$ 5.37 GV, and, 33.53 $-$ 36.12 GV, and, find that the lowest rigidity bin has the highest proton flux and the highest variation, $\sim$ 25 to 30\% from the average, with the solar cycle. We thus expect that cosmic rays detected by HST instruments will have kinetic energies consistent with the lowest rigidity of AMS reported in \citet{corti2019}. Using the relationships given in Appendix \ref{s:relativistic_relationships}, the corresponding kinetic energy range for the lowest rigidity bin is, $\sim$440 MeV to $\sim$553 MeV.

\vspace{-0.5cm}
\begin{table}[H]
    \centering
    \caption{The extracted frequencies for each instrument in Figure \ref{fig:lomb_scargle}.}
    \begin{tabular}{ccccc}
\toprule
Instrument & Peak 1 [cycles/day]   & Period [yr] & Peak 2 [cycles/day]  & Period [day] \\
\midrule
ACS/HRC    & 0.001577  & 1.74   & 0.02081 & 48.05            \\ 
ACS/WFC    & 0.000231 & 11.87  & 0.02089 & 47.87            \\ 
STIS/CCD   & 0.000209 & 13.10  & 0.02103 & 47.55            \\ 
WFPC2      & 0.000272 & 10.09   & 0.02064 & 48.46            \\ 
WFC3/UVIS  & 0.000237 & 11.54  & 0.02098 & 47.66            \\ 
\bottomrule
\end{tabular}
    \label{tab:periodogram}
\end{table}


\subsection{The South Atlantic Anomaly}\label{s:saa}
Since its discovery, the SAA has been, and will continue to be, an area of great interest as humanity moves to increases its presence in Low Earth Orbit. The SAA is characterized by an anomalously low value of 
the geomagnetic field intensity near the Earth's surface and drifts at a rate of  $0.36 \pm 0.06\degree$ W/yr  and $0.16 \pm 0.09\degree$ N/yr \citep{schaefer2016}. As a consequence of the weak intensity, particles trapped in the inner Van Allen radiation belts can more readily penetrate into the Earth's upper atmosphere \citep{heynderickx2002}.

HST routinely passes through the SAA \citep{lupie2002} whose boundary is defined by avoidance contours. Their coordinates are available via the \texttt{costools} Python package\footnote{\url{https://github.com/spacetelescope/costools}} maintained by the Cosmic Origins Spectrograph Instrument Branch at the Space Telescope Science Institute. To safeguard against damage to the electronics, astronomical observations are scheduled during SAA-free orbits. Occasionally, an HST calibration program is carried out to map the SAA boundary (\citet{barker2009}, \citet{martel2009}). Here we show the results of our analysis of HST Proposal 7061 for which dark exposures were made inside and outside the SAA.

Nineteen 60-second darks were taken through the SAA.  The position of HST during these observations is shown in Figure \ref{fig:hst_location} and the data set is listed in Table \ref{tab:stis_saa_estimates} in the Appendix. In Figure \ref{fig:stis_saa_img}, we show 150 by 150 pixel subsections of the darks to visually demonstrate the level of cosmic ray contamination in the SAA. 

\vspace{-0.2cm}
\begin{figure}[H]
\centering
    \includegraphics[scale=0.23]{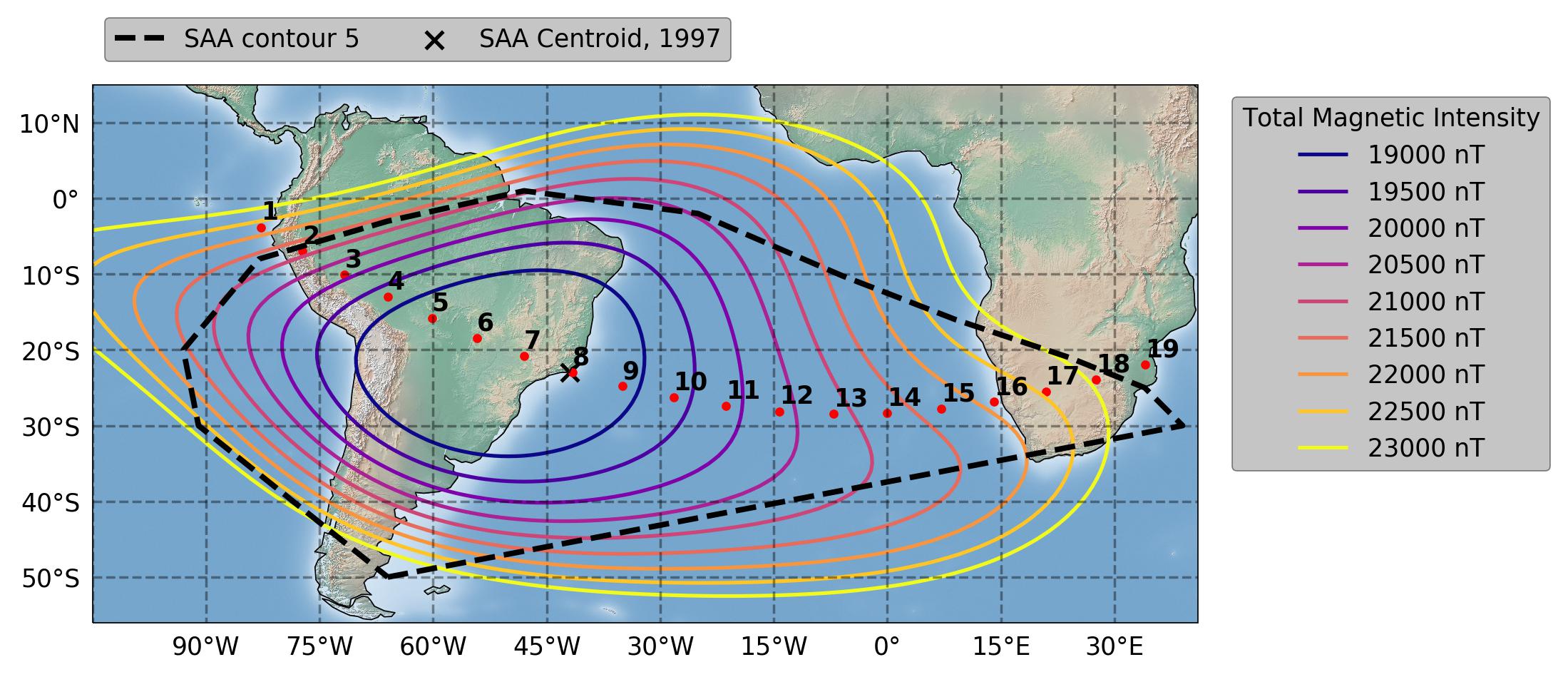}
    \caption{The location of HST during each of the 19 images taken as part of proposal 7061. The black, dashed polygon outlines the SAA region defined by SAA avoidance contour 5 (see \cite{lupie2002}). The contour lines correspond to the total magnetic intensity in 1997 at an altitude of 640 km in the vicinity of the SAA computed using the IGRF-13 model using \texttt{PmagPy} \citep{pmagpy}. Additionally, we show the contemporaneous centroid of the SAA reported by \citet{furst2009}. }
    \label{fig:hst_location}
\end{figure}
\vspace{-0.75cm}
\begin{figure}[H]
    \centering
    \includegraphics[scale=0.24]{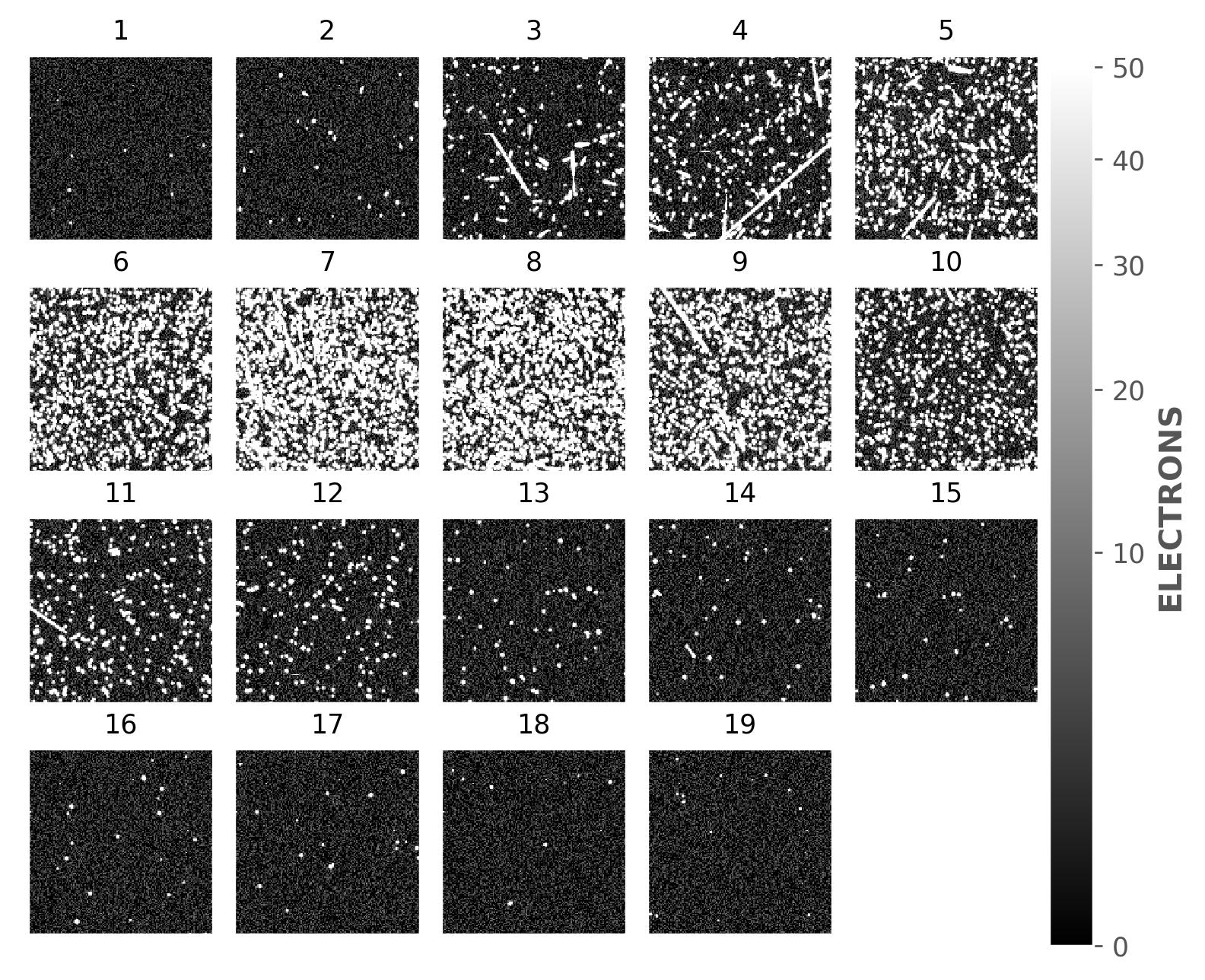}
    \caption{Each subplot shows a 150 by 150 pixel cutout from one of the 19 STIS images. The label at the top of each subplot indicates the image's number in the observing sequence displayed in Figure \ref{fig:hst_location}.}
    \label{fig:stis_saa_img}
\end{figure}

For exposures deep in the SAA, e.g. plots 5 to 10 in Figure \ref{fig:stis_saa_img}, it is clear that it is impossible to reliably extract statistics on individual cosmic rays due to the large number of overlapping cosmic rays. However, the STIS cosmic ray rejection algorithm does a superb at identifying all of the pixels affected by cosmic rays. Hence, by generating the label from the DQ array we can compute the total energy deposited by all cosmic rays and estimate the cosmic-ray particle flux as,

\begin{equation}
      \text{Number of Cosmic Rays} = \frac{E_{total}}{\langle E_{CR}\rangle * t * A}.
      \label{eq:rate_estimate}
\end{equation}
where 
\begin{itemize}
\item $E_{total} :=$ Total energy deposited by all cosmic rays in an image. [$e^-$]
\item $\langle E_{CR} \rangle :=$ Average energy deposited per image by all cosmic rays. [$e^-/s/cm^2$]
\item $t :=$ Total integration time for the image. [$s$]
\item $A :=$ Geometric area of the STIS/CCD (see Table \ref{tab: detectors}). [$cm^2$]
\end{itemize}

The value of $\langle E_{CR}\rangle$ can be determined in two ways using the data generated by our pipeline. The first method utilizes results presented in Tables \ref{tab:cr_statistics} and \ref{tab:cr_rate_statistics}. From Table \ref{tab:cr_statistics}, the average total energy deposited by a single cosmic ray is $\sim2621 e^-$. From Table \ref{tab:cr_rate_statistics}, the average cosmic-ray particle flux is 0.96 $\sim$1 particle $s^{-1}$ $cm^{-2}$. We compute $\langle E_{CR}\rangle$ as the product of these two measurements which yields a value of $\sim2516$ $e^{-} s^{-1}$ $cm^{-2})$. 

\begin{figure}[H]
    \centering
    \includegraphics[width=0.75\textwidth]{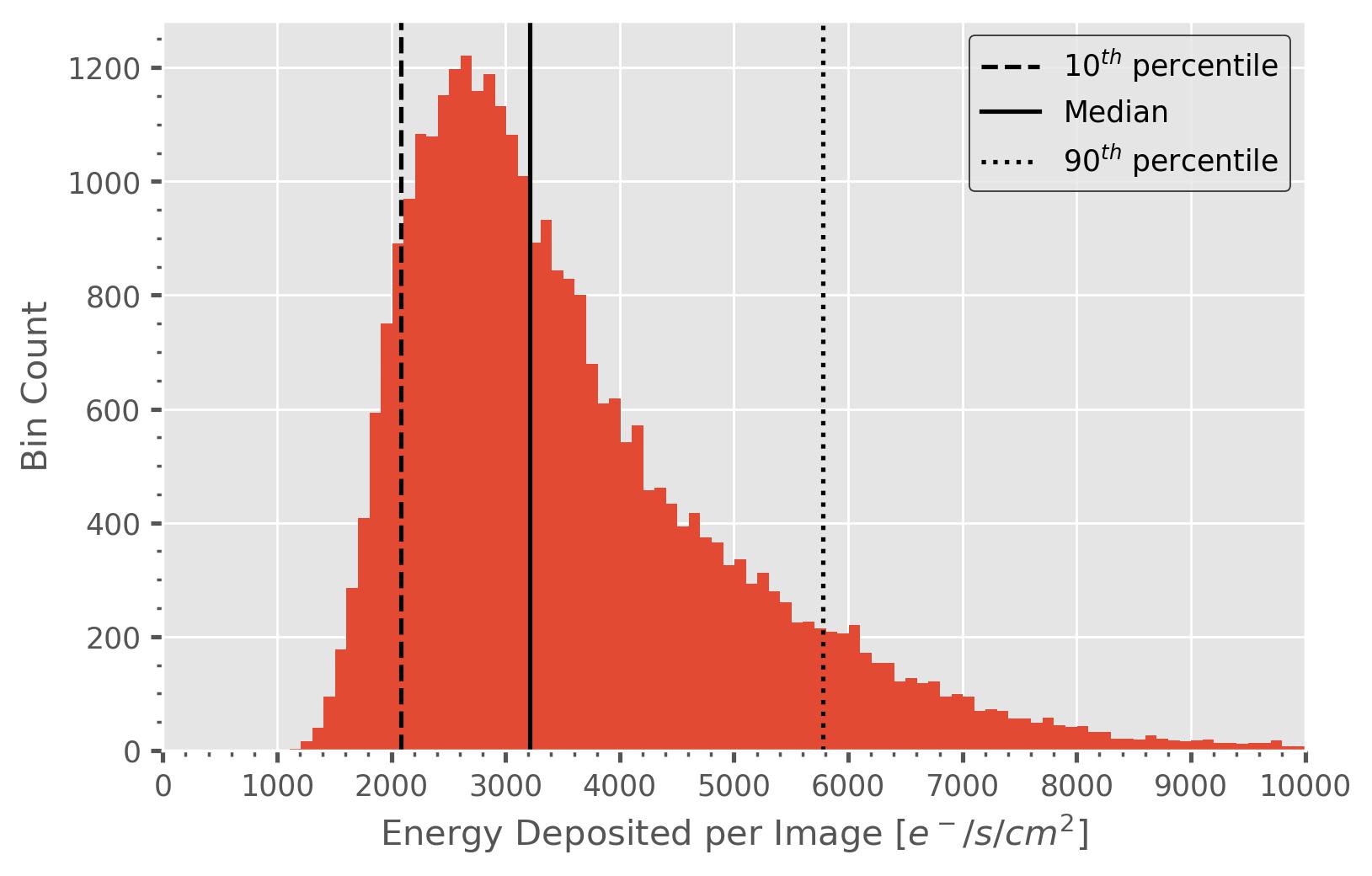}
    \caption{The distribution of the rate of energy deposition by cosmic rays in a single image for the STIS/CCD dataset. The dashed line at $2085e^{-}/s/cm^2$ marks the $10^{th}$ percentile, the solid line at $3215e^{-}/s/cm^2$ marks the median ($50^{th}$ percentile), and the dotted line at $5778e^{-}/s/cm^2$ marks the $90^{th}$ percentile.}
    \label{fig:total_energy_distribution}
\end{figure}

The second method utilizes the morphological parameters generated by our pipeline. The results for the cosmic rays identified in each image are stored as separate HDF5 datasets within a single HDF5 file. This allows us to derive a value for $E_{CR}$ from each image to generate a distribution. We compute $E_{CR}$ as the sum of the energy deposited by all of the identified cosmic rays, divided by the integration time and detector area. We analyze all STIS/CCD observations taken outside the SAA ($\sim31,000$) to generate the distribution in Figure \ref{fig:total_energy_distribution}. The median value is $\sim3215$ $e^{-}/s/cm^2$ and the most probable bin from the histogram is $~2657$ $e^{-}/s/cm^2$. Because of the positive skew, we use the most probable bin for the value of $\langle E_{CR}\rangle$ as it provides a more accurate estimate than the median.

 \vspace{-0.5cm}
The values derived using the two different methods agree to within 5\% and we adopt the value generated from second method as $\langle E_{CR}\rangle$. We use this value to calculate the cosmic-ray particle flux for each SAA image and list the results in Table \ref{tab:stis_saa_estimates} in the Appendix.  In Figure \ref{fig:stis_saa_total_energy}, we show the cosmic-ray particle flux as a function of the time elapsed since the first image in the observing sequence. The maximum cosmic-ray particle flux was $\sim$1100 particle $s^{-1}$ $cm^{-2}$ and it occurred in Observation 8 (o3st20gjq\_flt.fits) at a latitude of 22.96 $\degree$S, a longitude 41.61 $\degree$W. The minimum cosmic-ray particle flux was $\sim$1.4 particle $s^{-1}$ $cm^{-2}$ and it occurred in Observation 1 (o3st20gcq\_flt.fits) at a latitude of 3.81 $\degree$S, a longitude of 82.76 $\degree$W. These observations correspond to the maximum and minimum of the rate of energy deposition by cosmic rays of $\sim3\mathrm{x}10^{6}$ $e^-/s/cm^2$ and $\sim3570$ $e^-/s/cm^2$, respectively.

We find that the cosmic-ray particle flux increases by more than a factor of 800 from the edge to the center of the SAA. When the peak cosmic-ray particle flux was observed, HST was nearly co-located with the derived centroid of the SAA in 1997 reported by \citet{furst2009} (Figure \ref{fig:hst_location}) providing additional constraints on the morphology of the SAA at the time. 
\begin{figure}[H]
    \centering
    \includegraphics[width=0.85\textwidth]{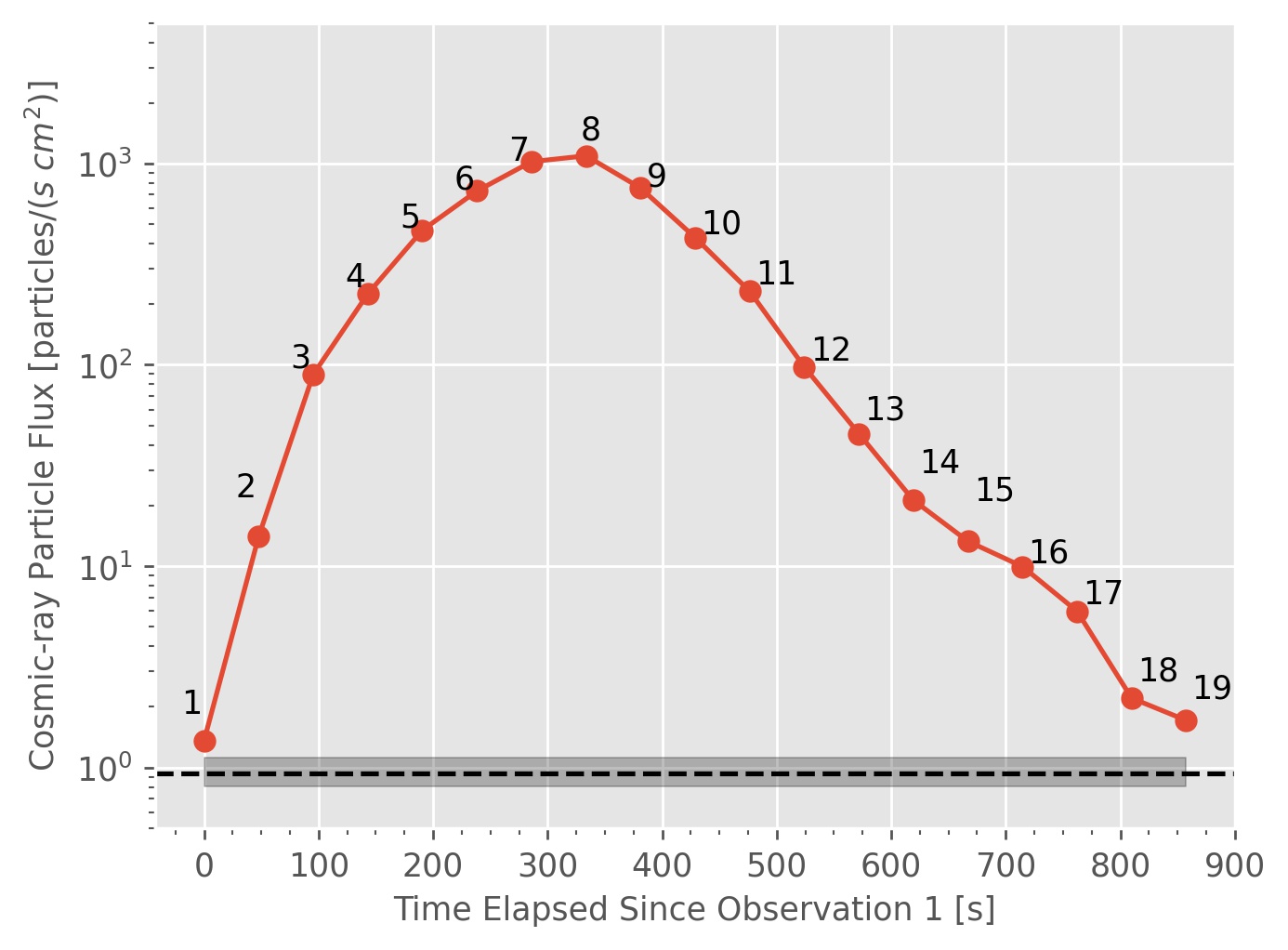}
    \caption{The cosmic-ray particle flux as a function of time elapsed since Observation 1 in Figure \ref{fig:hst_location}. Each point is labeled to indicate the observation used to make the measurement. The dashed horizontal line is mean value reported for the STIS/CCD in Table \ref{tab:cr_rate_statistics}. The gray shading denotes the interval bounded by the $25^{th}$ and $75^{th}$ percentiles of the cosmic-ray particle flux distribution shown in Figure \ref{fig:rate_hist}. }
    \label{fig:stis_saa_total_energy}
\end{figure}
\vspace{-0.75cm}
\subsection{Hot Spots}\label{s:hot_spots}
\begin{figure}[H]
    \centering
    \includegraphics[width=1.05\textwidth]{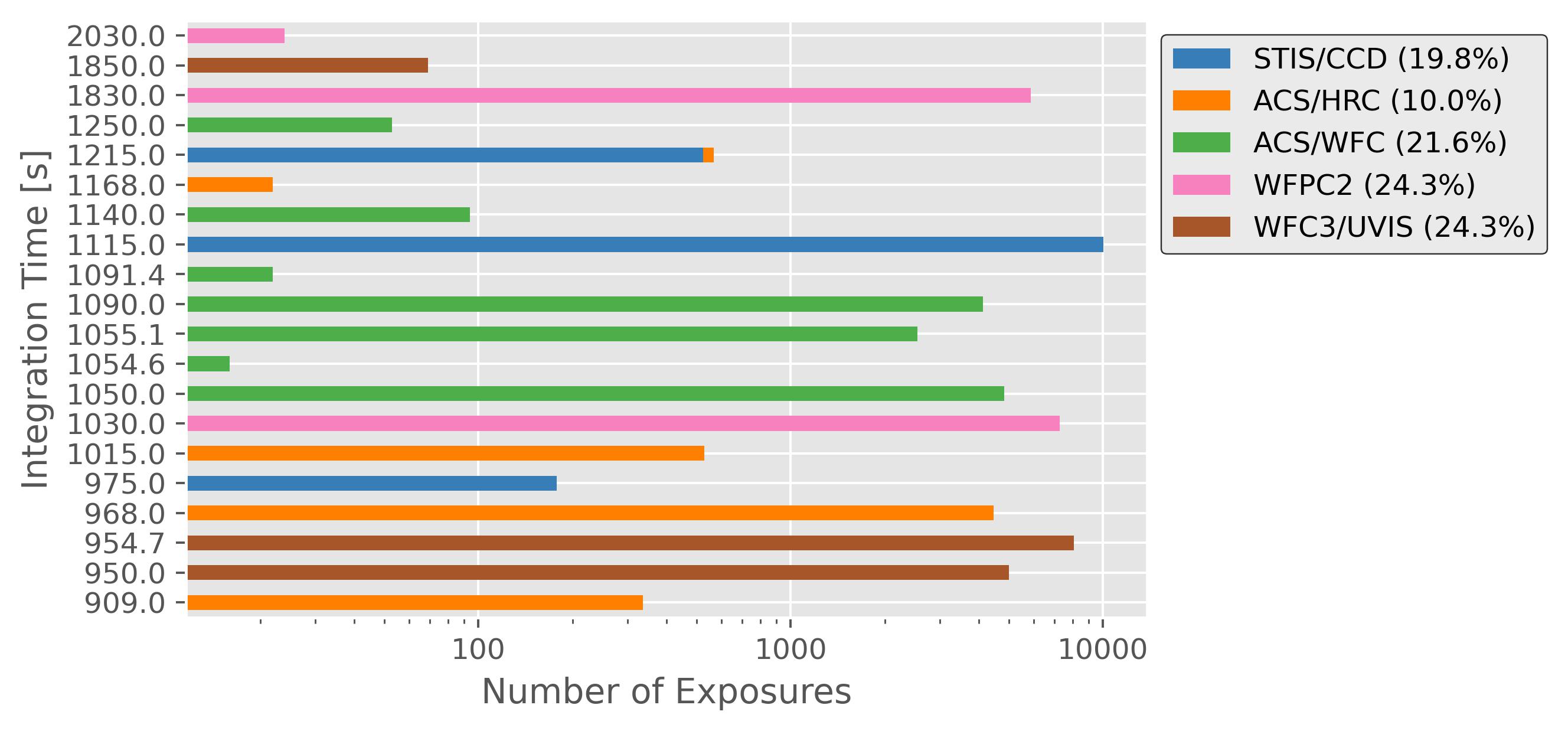}
    \caption{We analyze 54,215 images and compute the top 20 most frequently used integration times among any of 5 CCD imagers. Of the total integration time WFC3 and WFPC2 each contribute $\sim$24\%, STIS 19.8\%, and ACS (WFC+HRC), $\sim$31\%.}
    \label{fig:exptime_dsitribution}
\end{figure}

It is well known that the trajectory of charged particles are affected by the presence of external electric and magnetic fields. Since cosmic rays are  charged particles, we expect the cosmic-ray particle flux observed by HST to depend on the geomagnetic field.
We compare the cosmic-ray particle flux as a function of latitude and longitude to the total magnetic intensity at an altitude of 565 km (HST's 30-year average altitude). For this analysis we use cosmic-ray particle flux measurements from images where the integration time is greater than 800 seconds. In Section \ref{s:analyzing_crs}, we defined the integration time as the exposure time plus half the time required to readout the detector. When the exposure time and readout time are comparable, the fraction of  detected cosmic rays that impacted the detector during readout is no longer negligible. This amplifies any error introduce by computing the integration time as the sum of the exposure time and half the readout time. The readout times range between 29 seconds (STIS/CCD) to 120 seconds (WFPC2), and are listed in Table \ref{tab:readout}. Thus the choice of our 800-second cut is to minimize the error associated with cosmic rays striking during readout. In Figure \ref{fig:exptime_dsitribution}, we show the top 20 most common integration times that are longer than 800 seconds amongst the 5 imagers.
\vspace{-0.45cm}
\begin{table}[H]
\centering
    \caption{Full frame CCD readout times for each instrument}
    \begin{tabular}{l c l}
    \toprule
  CCD   & Readout Time & Reference \\
  & seconds  &\\
  \midrule
  WFC3/UVIS  & 96 & \citet{wfc3_ihb} \\
  ACS/WFC & 100 & \citet{mutchler_2005}, p. 3\\
  ACS/HRC & 26 & \citet{mutchler_2005}, p. 3\\ 
  STIS/CCD & 29 & \citet{stis_ihb}\\
  WFPC2  & $\begin{cases} 60 &\text{If exptime $\leq$ 180 s}\\ 120 &\text{If exptime $>$ 180 s} \end{cases}$ & \citet{wfpc2_ihb}\\
  \bottomrule
    \end{tabular}
\label{tab:readout}
\end{table}

\vspace{-1cm}
\vspace{-.45cm}

\begin{figure}[H]
\hspace{-1.2cm}
    \includegraphics[scale=0.9]{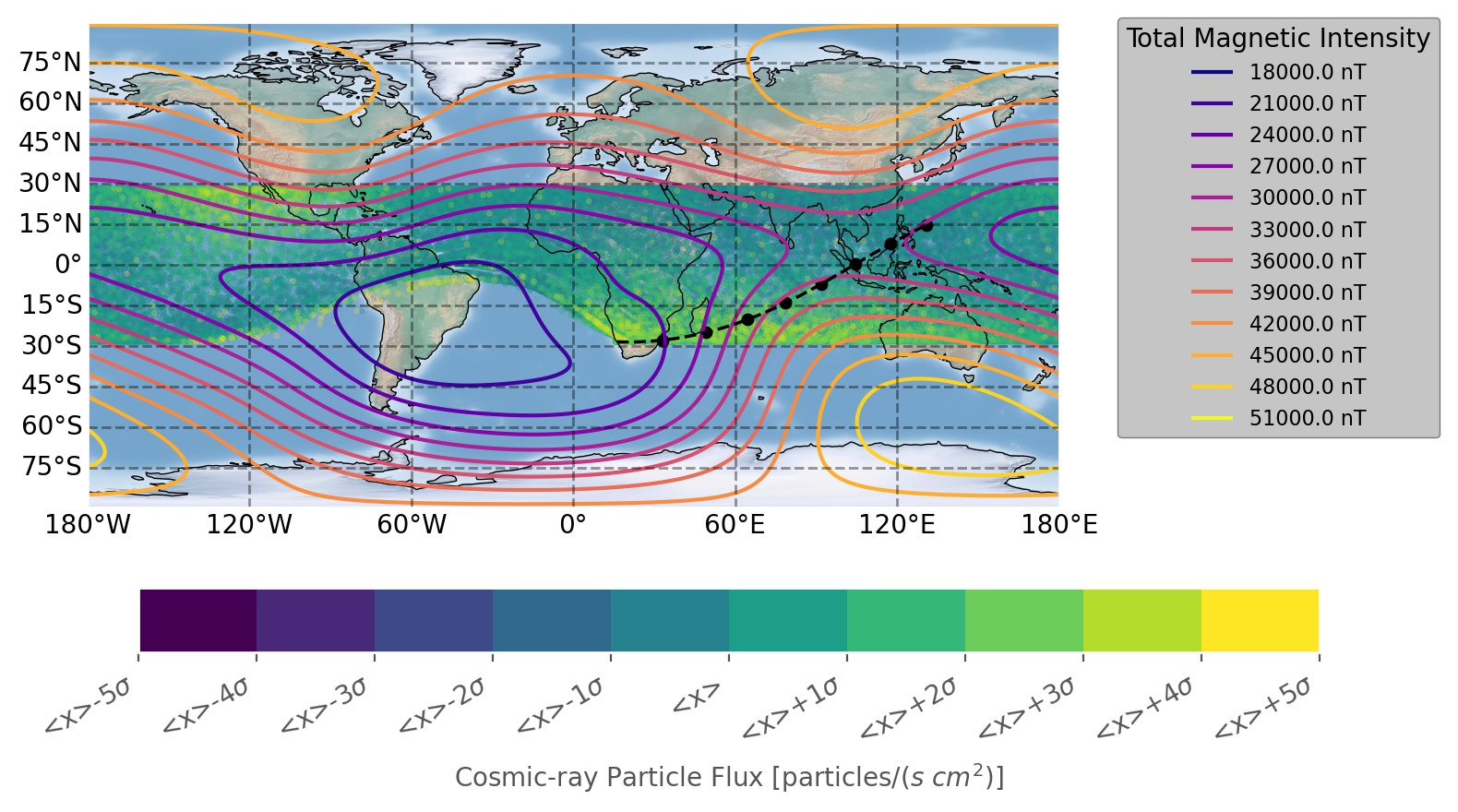}
    \caption{The observed cosmic-ray particle flux as a function of orbital position. Each point corresponds to a single observation where the integration was longer than 800 seconds. The color mapping corresponds to the observed cosmic-ray particle flux where the mean and standard deviation, $\langle x\rangle$ and $\sigma$, respectively, are sigma-clipped values computed using all 54,215 observations. The dashed, black line is an example of HST's ground track over a 2000 second exposure. The black points mark 250-second intervals. We assume an average date and altitude of 2005 and 565 km, respectively, for computing the total magnetic intensity using the IGRF-13 model. }
    \label{fig:rate_vs_location}
\end{figure}

After making this cut, there are 54,215 images distributed between the five CCD imagers with an average integration time of $\sim1119$ seconds. To compute the total magnetic field intensity, we use the IGRF-13 model for 2005 (the mid point of HST operational lifetime) and an altitude of 565 km, which approximately corresponds to HST's average orbital altitude. In Figure \ref{fig:rate_vs_location}, we superpose a plot of the distribution of the observed cosmic-ray particle flux as a function of HST's orbital position onto a map of the total magnetic field intensity.  

Each point in Figure \ref{fig:rate_vs_location} corresponds to the latitude and longitude at the start of the integration of a single dark frame, and is color coded (dark purple to bright yellow) according the to the observed cosmic-ray particle flux. The dashed, black line is an example of HST's ground track and the black points mark 250-second intervals along the path. Because of HST's orbital inclination, the projected latitude coverage is limited to $\pm$ 28.5\degree. Colored contour lines indicate the magnetic intensity, and range between 18000 nT (dark blue) to 51000 nT (yellow).

\vspace{-0.5cm}
Two ``hot spots" at the 5$\sigma$ level, are apparent. The northern region extends between 90\degree W and 150\degree W and 15\degree - 28.5\degree N. The southern area extends from the eastern edge of the SAA to Western Australia, i.e between 15\degree E to 120\degree E, and it north/south range is between 15\degree to 28.5\degree S.  These regions appear to correspond to locations of where the  magnetic field intensity is around 36000 nT. The average integration time of 1119 seconds corresponds to a track of 4 black points, so there is some smearing of the actual location and extent of each hot spot due the differences in the starting location of each observation. However, the robust sampling of latitude and longitude positions over the 25 year period minimizes the impacts of this because there are just as many observations taken while entering each hot spot as there are while exiting.

\subsection{Kinetic Energy Estimation}\label{s:energy_estimation}
In order to compute an estimate of the average kinetic energy of the observed cosmic rays, we make a series of simplifying assumptions. 

Firstly, we assume that the detected cosmic rays are relativistic protons with sufficiently high kinetic energy to penetrate down through the magnetosphere to HST's orbit: 
low altitude of $\sim$540 km (see Figure \ref{fig:altitude}) and an orbital inclination of $\pm$ 28$\degree$.  At this orbital inclination, the vertical geomagnetic cutoff rigidity at an altitude of 450 km is $\geq$ $\sim$ 5 GV (Figure 7 in \citet{smart2005}) corresponding to a kinetic energy of approximately 4 GeV or higher (i.e. $\beta\geq 0.98$), for high-energy protons originating outside the magnetosphere.

Secondly, we ignore shielding losses for arbitrary trajectories through the HST observatory. Hence the kinetic energy estimated here establishes a \emph{lower limit} for the typical cosmic ray interacting with the CCD imagers. Note that the presence of trapped radiation in low earth orbit means that this is not a hard limit.

\begin{figure}[H]
    \centering
    \includegraphics[scale=0.8]{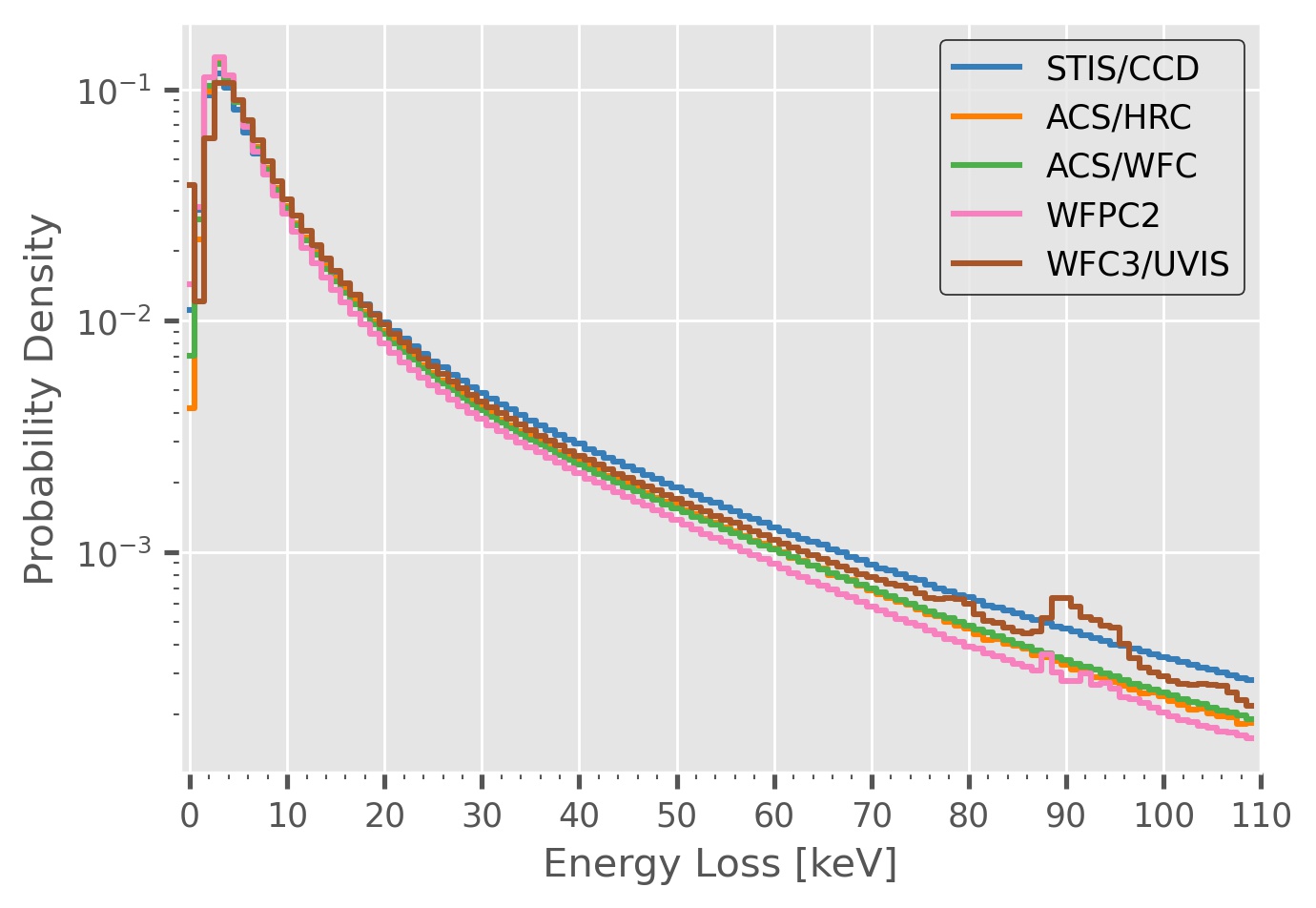}
    \caption{The observed energy loss distributions for each instrument. For energy losses below $\sim20$ keV, the distributions are very similar. After $\sim20$ keV, the distributions begin to diverge from one another due to difference in shielding and instrument location within the observatory. }
    \label{fig:energy_loss_full}
\end{figure}

Finally, to determine the energy deposited by cosmic rays we convert the detected signal (in electrons) to energy (eV).  The average operating temperature of the CCDs is  $-82$ $\degree$C (\citet{acs_ihb}, \citet{stis_ihb}, \citet{wfc3_ihb}).  At this temperature, the average energy required to produce an electron-hole pair in silicon is $\sim$ 3.71 eV \citep{lowe2007}. Thus, the energy deposited in eV is $\text{(3.71 eV/electron)} \times \text{(number of electrons)}$. The distribution of energy deposited, equivalent to cosmic ray energy loss, for the five CCD imagers is shown in Figure \ref{fig:energy_loss_full}. 

\subsubsection{Estimating the Energy-loss Probability Density Function}

For the subsequent analyses we use a subset of the data covering the time period between 2001 and 2005 when four CCD imagers operated simultaneously on HST:  ACS/HRC, ACS/WFC, STIS/CCD, and WFPC2 (Table \ref{tab: detectors}). Further, we restrict our analysis to cosmic rays with path lengths between 280 and 300 $\mu$m, primarily because this allows us to make a direct comparison to previous laboratory results. \citet{bichsel1988} report results for protons with relativistic kinetic energies of 1.2 GeV and 7.1 GeV (corresponding to $\beta\gamma$ equal to 2.1 and 8.5 respectively) passing through 290 $\mu$m of silicon. The data presented in Table 9 of \citet{bichsel1988} are a reproduction of the results obtained by \citet{bak1987}.

\begin{figure}[H]
    \centering
    \includegraphics[scale=1.05]{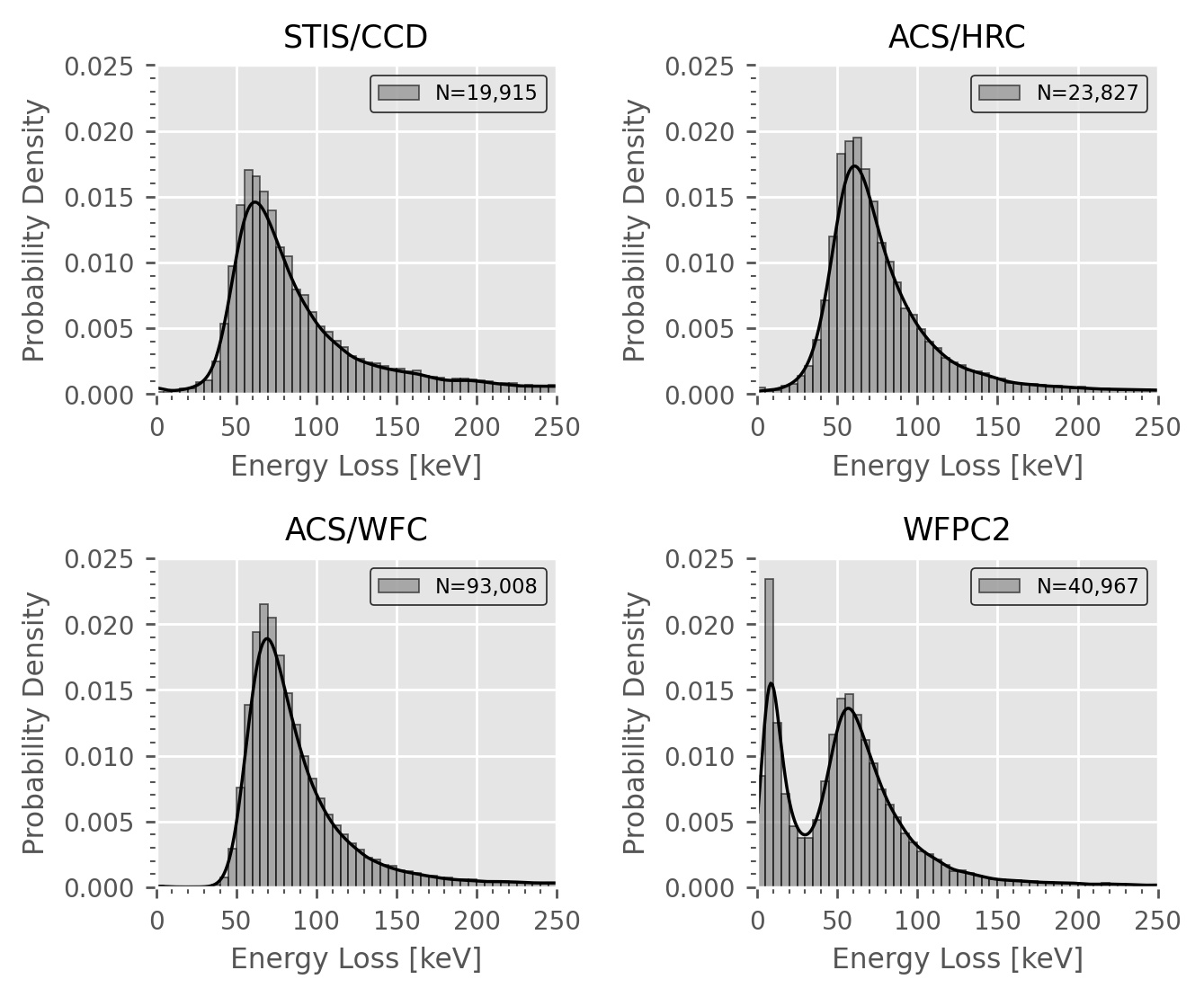}
    \caption{The energy-loss distributions for the 4 imagers. The solid, black line is the KDE-derived probability density function. The gray shading is a histogram of the dataset generated using a bin size of 5 keV. Each histogram has been normalized to yield a probability density function.}
    \label{fig:energy_loss_distribution}
\end{figure}

We calculate the probability density function of the energy loss distribution using the \texttt{statsmodels} \citep{seabold2010} Python package to perform the kernel density estimation, and estimate an optimal bandwidth via a maximum-likelihood cross-validation technique.  Figure \ref{fig:energy_loss_distribution} shows a histogram of the  energy-loss distribution with bin size of 5 keV (gray rectangles) for track lengths between 280 and 300 $\mu$m, and the corresponding KDE-derived probability density function (solid, black line) for each instrument. The thinned, backside illuminated CCDs (ACS/HRC, ACS/WFC, and STIS/CCD) all have a single peak at $\sim$60 keV. The thick, frontside-illuminated CCD (WFPC2) has two clear peaks. The first is around $\sim$5 keV and the second is around $\sim$55 keV. 

The bimodal energy-loss distribution for WFPC2 is a consequence of its CCD structure. Front-illuminated CCDs that are not thinned can have thick, up to $\sim 500 \mu$m, bulk silicon substrates even though the thickness of the photosensitive (epitaxial) layer is comparable to those of thinned,  back-illuminated detectors at $\sim 10-20$ $\mu$m. Cosmic rays deposit energy throughout their path through the detector.  Electrons generated in or near the depletion region are more likely to be collected, whereas those generated in the substrate and field free regions are more likely to be lost to recombination (partial events). Additionally, split events can occur when charge diffuses from one pixel to another.  Thick CCDs have a higher fraction of partial and split events compared to thinned CCDs which results in a lower charge collection efficiency \citep{janesick2001}.

DQ labeling and threshold labeling (Section \ref{s:cr_algorithm}) minimize the effects of split events by identifying all pixels directly affected by each cosmic ray, as well as their neighbors. The calibration of partial events is much more difficult and requires knowledge of the depth of the interaction.  Due to its bimodal distribution, we omit WFPC2 from the remainder of the kinetic energy estimation analysis.

\subsubsection{The Landau Distribution}

Ionization due to inelastic collisions between an incoming particle and the electrons in the silicon substrate is the principal energy loss mechanism in solid-state detectors. For relativistic charged particles interacting with matter, the maximum possible energy transfer in a single collision between a particle of mass M and an electron is given by \citep{pdg2018},
\begin{equation}
    W_{max} = \frac{2 m_e c^2 \beta^2 \gamma^2}{1+2\left(\gamma m_e/M\right) + \left( m_e/M\right)^2},
    \label{eq:max_loss}
\end{equation}
\noindent{where $m_e$ is the mass of an electron, $\beta=v/c$, and $\gamma=\sqrt{1/(1-\beta^2)}$.
The energy loss of the  incident particle is subject to random fluctuations arising from the probabilistic nature of the collisions.}

For the case when the energy loss is small compared to the particle's kinetic energy (i.e. thin detectors) and the electron binding energy is negligible, the energy loss distribution is described by  the Landau distribution \citep{landau1944} where the peak value corresponds to the most probable energy loss:
\begin{equation}
    {}_L\Delta_p = \xi \left[ \ln{\frac{2m_ec^2\beta^2\gamma^2}{I} + \ln{\frac{\xi}{I}}} + 0.200 - \beta^2 -\delta(\beta\gamma)\right],
    \label{eq:landau}
\end{equation}
\noindent{\textit{I} is the mean excitation energy of the material, and $\xi$ is the Landau parameter describing the typical energy loss for the material. Restating the conditions under which the Landau distribution describes energy loss:
\begin{itemize}
    \item $k = \xi/W_{max} \ll 1$ and 
    \item $\xi \gg E_\text{bind}$, where $E_\text{bind}$ is the electron binding energy
\end{itemize}}
\noindent{When the electronic binding energy is no longer negligible, the distribution is more adequately described by the convolution of the Landau distribution with a Gaussian (\citet{bak1987}, \citet{bichsel1988}, \citet{meroli2011b}}

For silicon, the value of $\xi$ is
\begin{equation}
     \xi = 0.017825*t/\beta^2 \text{ keV/$\mu$m},
     \label{eq:landau_parameter}
\end{equation}
\noindent{and $t$ is the thickness of the detector  (\citet{bak1987}; \citet{bichsel1988}). 
The FWHM of the distribution is approximately $w_L = 4.016\xi$.   Silicon's maximum electronic binding energy occurs in the K-shell, $E_\text{K}=1.84$ keV. By substituting $E_{\text{K}}$ for $E_\text{bind}$ for the condition $\xi \gg E_\text{K}$, we obtain,}
\begin{equation}
    t/\beta^2 \gg 103 \text{ $\mu$m}.
    \label{eq:landau_bind}
\end{equation}

We compute $W_{max}$ (Eqn. \ref{eq:max_loss}) and the Landau parameter,  $\xi$,  (Eqn. \ref{eq:landau_parameter}) for the HST data assuming cosmic rays are protons with kinetic energies $\ge$ 4 GeV, ($\beta = 0.982$), and, for the Bichsel results (See Table \ref{tab:landaunumbers}).   From these quantities we see that the first condition for the Landau distribution, $k=\xi/W_{max} \approx 1\mathrm{x}10^{-4} \ll 1$, is satisfied, but the second, $t/\beta^2 \gg 103 \mu$m, is not, indicating that the energy loss distribution is best described by the convolution of a Gaussian with the Landau distribution.

\begin{table}[H]
    \centering
    \caption{The Landau parameters computed in this work and \citet{bichsel1988}.}
    \begin{tabular}{c c c c c c c}
    \toprule
        path length [$\mu$m] &$\beta$ & $\xi$ [keV] & W$_\text{max}$ [keV] & $\xi$/Wmax & t/$\beta^2$ [$\mu$m] & Source \\
        \midrule
        290 & 0.903 & 6.339 & 4.496$\times10^3$ & 1.42$\times10^{-3}$ & 356 &\citet{bichsel1988}\\
        290 & 0.982  & 5.360 & 2.746$\times10^4$ & 1.95$\times10^{-4}$ & 301 & this work\\
        290 & 0.993 & 5.242 & 7.316$\times10^4$ & 7.17$\times10^{-5}$ & 294 & \citet{bichsel1988} \\
        \hline
    \end{tabular}
    \label{tab:landaunumbers}
\end{table}


Using the Python implementation of \texttt{Minuit2} \citep{hatlo2005}, \texttt{iminuit} \citep{dembinski2020}, we fit a Landau convolved with a Gaussian to the kernel density estimation of the underlying probability distribution for the energy losses. In each fit, the 
free parameters are: the most probable energy loss of the convolved distribution ($\Delta_{p}$), the width of the Landau distribution ($\xi$), the width of the Gaussian distribution ($\sigma_{\text{Gauss}}$), and the height of the convolved distribution ($A$). We compute the RMS error as, 
\begin{equation}
    \sigma_{RMS} = \sqrt{\Big \langle (y_{\text{fit}} - y_{\text{kde}})^2 \Big \rangle}.
    \label{eq:rms_err}
\end{equation}
and the normalized RMS error using the interquartile range ($IQR$),
\begin{equation}
    \sigma_{NRMS} = \sqrt{\bigg\langle \left(\frac{y_{\text{fit}} - y_{\text{kde}}}{IQR_\text{kde}}\right)^2\bigg\rangle}.
    \label{eq:nrms_err}
\end{equation}

\begin{figure}[H]
    \centering
    \includegraphics[scale=0.95]{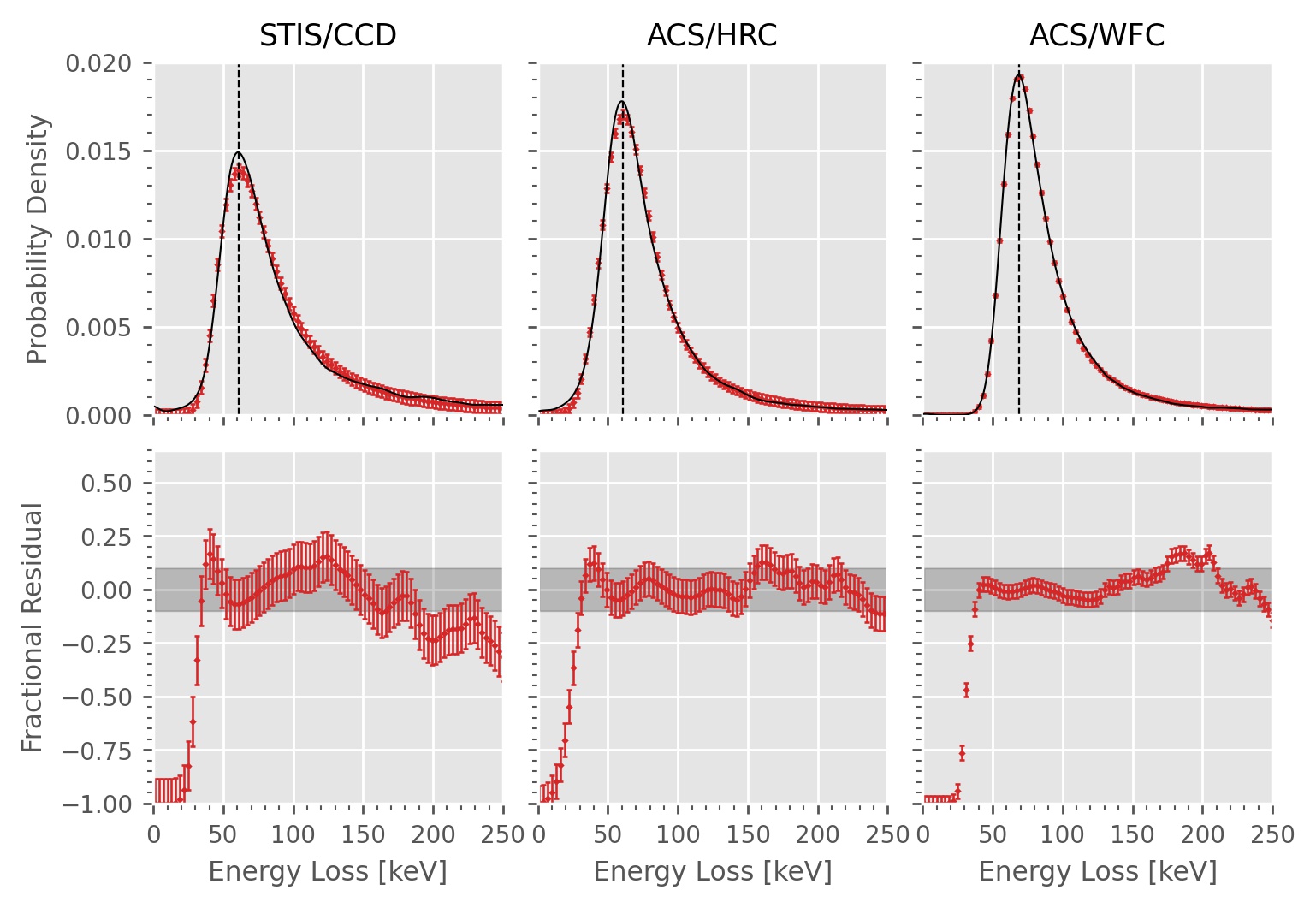}
    \caption{Top Row: The best-fit distribution is shown with red datapoints and the error bars correspond to $1*\sigma_{RMS}$. The solid, black line is the KDE-derived probability density function. Bottom Row: The fractional residuals for each fit computed as $(y_{fit} - y_{kde})/y_{kde}$. The error bars correspond to $1*\sigma_{NRMS}$ and the grading shading indicates $\pm 10\%$.}
    \label{fig:langau_fit}
\end{figure}

Figure \ref{fig:langau_fit} shows the best-fit distributions and their fractional residuals. In Table \ref{tab:langau_fit}, we report the best-fit parameters, the 68\% confidence intervals, and the normalized RMS error. When the energy losses are small ($< 20$ keV) the fits are poor.  
However the overall shape and peak of the distribution are fit  well.  

\begin{table}[H]
    \centering
    \caption{Best-fit parameters for the three CCD imagers analyzed. For each parameter, the 68\% confidence interval generated using the MINOS algorithm \citep{james1975} is given in parentheses. }
    \begin{tabular}{cccccccccc}
    \toprule
        {}& \multicolumn{2}{c}{$\Delta_{p}$ [keV]} & \multicolumn{2}{c}{$\xi$ [keV]} &
        \multicolumn{2}{c}{$\sigma_{\text{Gauss}}$ [keV]} & $\sigma_{NRMS} [\%]$\\
        \midrule
         ACS/HRC & \multicolumn{2}{c}{60.86, (60.72, 61.00)} & \multicolumn{2}{c}{ 7.98, (7.92, 8.04)} & \multicolumn{2}{c}{10.35, (10.17, 10.52)} & 7.97 \\
         
         ACS/WFC & \multicolumn{2}{c}{68.89, (68.81, 68.98)} & \multicolumn{2}{c}{7.91, (7.86, 7.95)} & \multicolumn{2}{c}{6.59, (6.50, 6.68)} & 3.33 \\
         
         STIS/CCD & \multicolumn{2}{c}{61.11, (60.94, 61.28)} & \multicolumn{2}{c}{12.13, (12.02, 12.23)} & \multicolumn{2}{c}{4.11, (3.71, 4.48)} & 11.51 \\
    \end{tabular}
    \label{tab:langau_fit}
\end{table}

\begin{figure}[H]
    \centering
    \includegraphics[scale=0.85]{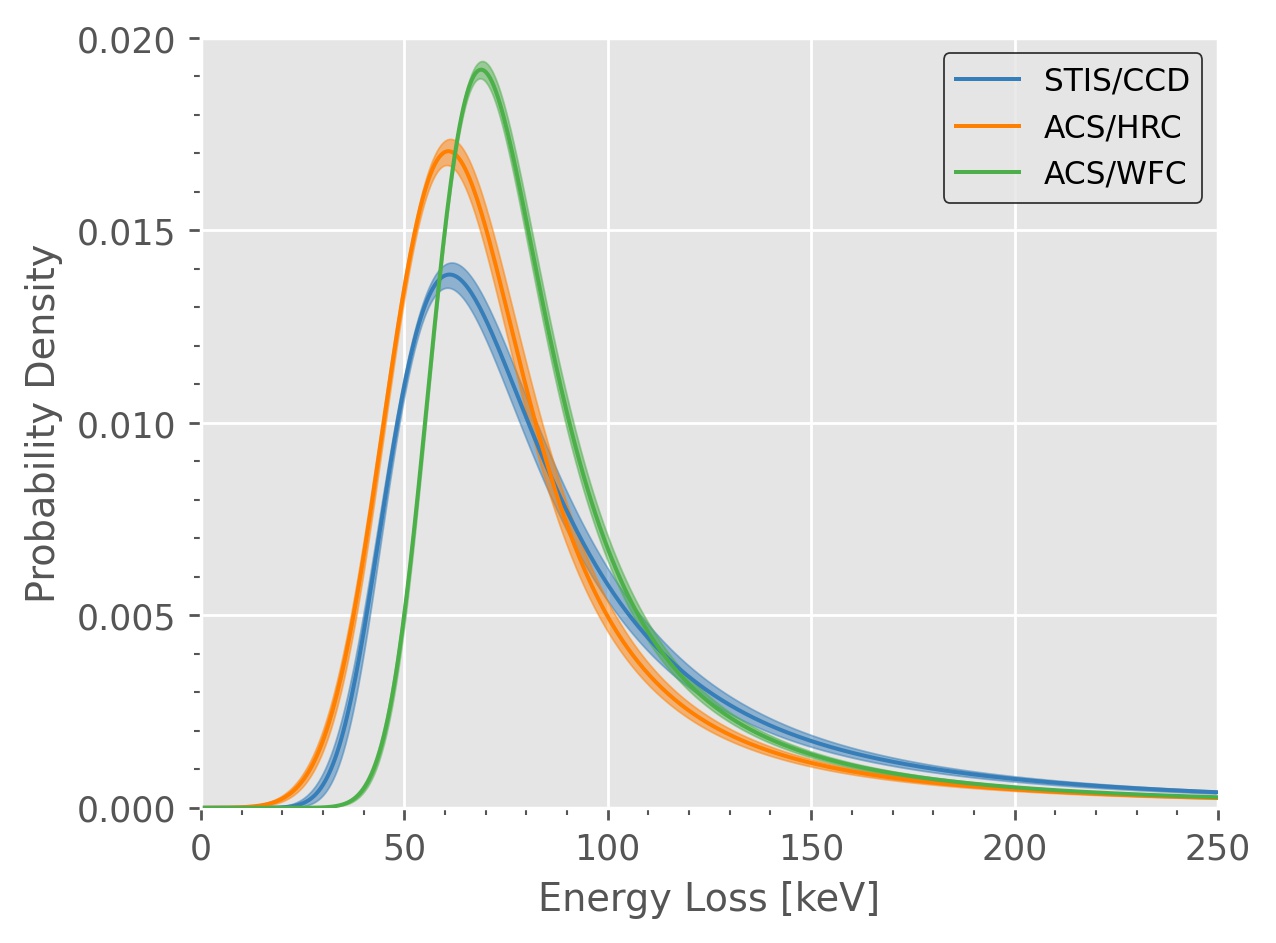}
    \caption{The best-fit straggling distributions with 3$\sigma$ error bands overlaid.}
    \label{fig:langau_fit_comparisons}
\end{figure}

Figure \ref{fig:langau_fit_comparisons} shows the best-fit distributions and their corresponding 3$\sigma$ error bands generated via MINOS. The peak of the ACS/WFC distribution is offset from the ACS/HRC and STIS/CCD distributions by $\sim 8$ keV. After losses exceed  $\sim$150 keV, the ACS/HRC and ACS/WFC distributions converge to within $\sim 8\%$ of one another. In the same regime, the STIS/CCD distribution is systematically higher than both ACS/HRC and ACS/WFC by an average of $\sim 50\%$. We measure the FWHM, $w$, of each distribution and find widths of 40.27 keV, 37.05 keV, and 44.27 keV for ACS/HRC, ACS/WFC, and STIS/CCD respectively.

The best fit is for ACS/WFC where the normalized RMS error is 3.33\%; the worst fit is for STIS/CCD where the normalized RMS error is 11.51\%. The poor quality of the STIS/CCD fit is driven by the lack of thermal control of the CCD. In May 2001, the STIS Side-1 electronics box failed and the instrument began operation using the Side-2 electronics, which is not able to precisely control the operating temperature of the CCD.  The minimum energy  required to produce an electron-hole pair is temperature dependent, the change in energy per degree is  -0.013\%/K \citep{lowe2007}. At warmer temperatures, the same cosmic ray generates more electrons than at colder temperatures.


In Table \ref{tab:bichsel_comparison} we list our derived  values of $\Delta_p$ and $w$ and, for comparison, the values reported by \citet{bichsel1988} for protons with kinetic energies of 1.2 GeV and 7.1 GeV traversing 290 $\mu$m of silicon. The most probable energy losses calculated for the HST CCDs are lower by $\sim$20\%  and the widths are larger by 36\% compared to  \citet{bichsel1988}. 

The differences are due to the particle populations under investigation. Primary cosmic rays are a mix of 89\% protons, 9\% helium nuclei, 1\% heavy nuclei and 1\% electrons, and have a broad energy distribution with a peak of approximately 0.3 GeV.  \citet{bichsel1988}'s values are for data taken in a laboratory setting for protons with fixed kinetic energies.



\begin{table}[H]
    \centering
        \caption{The most probable energy losses and FWHMs estimated from this work (ACS/HRC, ACS/WFC, STIS/CCD) and those reported in \citet{bichsel1988} for 1.2 GeV and 7.1 GeV protons traversing 290 $\mu$m of silicon.}
    \begin{tabular}{cccccc}
        \toprule
         {} & ACS/HRC & ACS/WFC & STIS/CCD &
         1.2 GeV & 7.1 GeV \\
        \midrule
         $\Delta_{p} [keV]$ & 60.86 & 68.89 & 61.11 & 82.46 & 76.91 \\
         $w [keV]$ & 40.27 & 37.05 & 44.27 & 31.01& 28.53\\
         \bottomrule
    \end{tabular}
    \label{tab:bichsel_comparison}
    
\end{table}

We do not expect excellent agreement between our results and \citet{bichsel1988}'s.  Rather,  the goal of the comparison is to ensure our results are not orders of magnitude off from theoretical models and their high-precision experimental tests.

\subsubsection{Computing the Kinetic Energy}
Using the relationship between $\xi$, $\beta$, and $t$ defined by Equation \ref{eq:landau_parameter}, we use the value of $\xi$ obtained from the fit and the path length (280 $\mu$m $\leq$ t $\leq$ 300 $\mu$m) to solve for $\beta$, 

\begin{equation}
    \beta(t) = \sqrt{\frac{0.017825 * t}{\xi_{\text{fit}}}}
\end{equation}

and compute the relativistic kinetic energy, $T = (\gamma - 1)*E_o$, where $E_o$ is the rest energy of the proton (Table \ref{tab:beta}).

\begin{table}[H]
\centering
\caption{Kinetic energy of cosmic rays with path length of $\sim$290 $\mu$m}
\begin{tabular}{ccccc}
\toprule
Instrument & $\beta_{\text{280$\mu$m}}$ & T($\beta_{\text{280$\mu$m}}$) [MeV] &$\beta_{\text{300$\mu$m}}$ & T($\beta_{\text{300$\mu$m}}$) [MeV] \\
\midrule
ACS/HRC  &  0.791 & 594.23 & 0.818 & 694.59 \\
ACS/WFC  & 0.794  & 606.44 & 0.822  & 710.45 \\
STIS/CCD    &  0.641 & 284.48 &  0.664 &  316.17\\

\bottomrule
\end{tabular} 
\label{tab:beta}
\end{table}
\noindent{The average kinetic energy of particles leaving  tracks between 280 and 300 $\mu$m is 534 $\pm$ 117 MeV. For comparison, the minimum energy for a proton to be stopped in 10 $\mu$m of Silicon is 5 MeV indicating that the vast majority of these cosmic rays pass right through the CCDs analyzed. }

\subsubsection{Discussion}

\begin{figure}[H]
\centering
\includegraphics[scale=0.91]{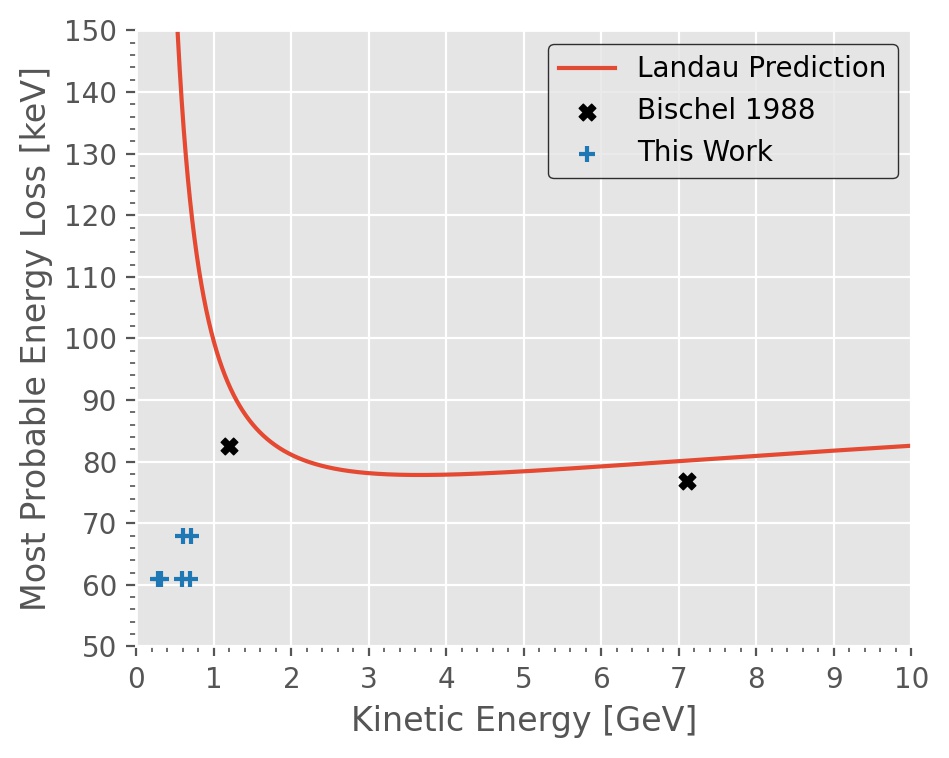}
\caption{The most probable energy loss for a proton traversing 290 $\mu$m of silicon. The black ``x" denotes the values presented in Table 9 of \citet{bichsel1988}. The blue 
``+" denotes the values obtained in this work. The solid, red line is the value of the peak of the Landau distribution computed using Equation \ref{eq:landau} assuming protons and a silicon target with a thickness of 290 $\mu$m. }
\label{fig:comparison_with_landau}
\end{figure}

In the preceding section we determined the relativistic speed, $\beta \sim 0.8$, of particles leaving $\sim$ 290 $\mu$m tracks, and thus their average kinetic energy, $\sim 0.5$ GeV. Particle kinetic energy versus the most probable energy loss is plotted in Figure  \ref{fig:comparison_with_landau} for a proton traversing 290 $\mu$m of silicon.  Shown in the figure are the Landau prediction (Equation \ref{eq:landau}), the results from \citet{bichsel1988}, and the results from this work. The energy losses for particles with kinetic energy larger than approximately 1.7 GeV are degenerate, i.e. two particles with distinct kinetic energies can have identical values for the most probable energy loss.

At 7.1 GeV, the difference between the Landau prediction and \citet{bichsel1988} is 5\%, but at  1.2 GeV, the prediction is almost 20\% larger than the measured value.   However, for our estimated average kinetic energy of  $\sim 0.5$ GeV, the most probable energy loss  predicted by the Landau distribution is $\sim 2$ times larger. Our measured most probable values are 20\% lower than the 1.2 GeV lab measurements. There are two dominant factors that contribute to this apparent discrepancy.

First, HST is exposed to cosmic rays with a broad range of kinetic energies. The population of particles comprising the cosmic rays that impact HST includes low energy trapped radiation,  high energy solar-energetic particles, and high energy galactic cosmic rays. From Equations \ref{eq:landau} and \ref{eq:landau_parameter} we see that the most probable energy is proportional to 1/$\beta^2$. Thus cosmic rays with lower kinetic energies will deposit more energy compared to cosmic rays with higher kinetic energies. The inability of HST to determine the incident kinetic energy of each detected cosmic ray means that each particle is treated with equal weight in our determination of the most probable energy loss. This results in a broadening of the energy loss distribution (larger $\xi$) due to contributions from lower energy particles which leads to a smaller estimation for $\beta$ through Equation \ref{eq:landau_parameter}. 


Second, the CCDs on HST do not have charge collection efficiencies of 100\%. This means that some of the electrons generated as a result of the interaction with the cosmic ray are lost. While this effect is more pronounced in the thick, frontside-illuminated CCDs, it is present in both types \citep{janesick2001}. This effect is dependent on the depth of the interaction within the pixel; the farther from the gate the lower the CCE. \citet{janesick2001} and \citet{Meroli2011a} examined the CCE of CCD and CMOS detectors, respectively, finding typical values of $\sim 0.8$. The imperfect CCE manifests as a systematic underestimation of the most probable energy loss.
While the effects of an imperfect CCE can be minimized by analyzing specific pixels along the particle track \citep{Meroli2011a}, the inability to determine the incident kinetic energy of each individual cosmic ray cannot be overcome. 

The net result of these two factors is the apparent low, kinetic energy of the incident cosmic ray, and, small value of the most probable energy loss.  A full treatment of these effects is beyond the scope of this paper, and reserved for future analysis.  To further evaluate the validity of our approach, we compare the estimation obtained here with additional works.

\citet{didkovsky2006} (hereafter D06) estimated the range of kinetic energy of incident particles by modelling the proton energy losses deposited in single-pixel cosmic ray events in the SOHO/EIT CCD. Modelled energies range from $\sim$30 MeV to $\sim$500 MeV. Energies are converted to differential proton fluxes and compared against data from the Geostationary Operational Environmental Satellite (GOES) Energetic Particle Sensor (EPS). For the two SEP (solar energetic particle ) events D06 examined, they find good correlation between their derived differential proton flux and GOES'. This method works well for the single-pixel events examined by D06 because the constrained angle of incidence (Figure \ref{fig:cr_trajectory}) allows isolating distinct energy ranges corresponding to specific CCD signal ranges.

 
\citet{shen2016} (hereafter SQ16) analyzed multi- and single-pixel CR events. They compared the extracted CR count rates to GOES data by analyzing measurements obtained from 23 different solar proton events (SPEs). They found the best correlation to be with the P6 channel on GOES 11, which probes energies between 80 and 165 MeV. They performed a similar correlation analysis with 39 events that were also observed by the SOHO Energetic and Relativistic Nuclei and Electron (ERNE) experiment. The highest correlation was with SOHO/ERNE channel 3, which probes energies between 118 and 140 MeV. By removing short-term fluctuations from SPEs, they analyzed the slowly varying, low-level background due to GCRs. To convert their GCR count rates to physical units, they utilize similarly smoothed SOHO/ERNE data to derive a scaling factor between the two datasets (Figure 8 in SQ16).

While both these methods provide estimations of the differential proton flux, the differences between the SOHO and HST mission designs reduces their applicability to our work. As a solar observatory, the telescope pointing is always fixed on the Sun. The SOHO/EIT CCD is at the rear end of the spacecraft and it is centered on the optical axis (see Figure 2 in \citet{soho_eit}). The CCD is oriented such that the surface normal is pointing radially towards the Sun. Hence the CCD surface is perpendicular to the direction of the particle flux during an SEP event which maximizes the number of single-pixel events during an SEP event. Additionally, HST does not have an in situ particle detector capable of accurately identifying particle types and energies which complicates the derivation of an accurate scaling factor. Next, HST's pointing is not fixed on a single target but changes constantly as it observes astrophysical sources, and when HST is on Earth's sunlit side, the pointing is unconstrained. Finally, HST is orbiting in low earth orbit inside the Earth's magnetosphere, whereas SOHO is in a halo orbit at L1 outside the Earth's magnetosphere.

Instead we compare our results with those derived using observations from the Alpha Magnetic Spectrometer (AMS) experiment on board the International Space Station (ISS). The ISS has a similar orbital altitude as HST, 450 km vs. 540 km, though a slightly steeper orbital inclination of  51$\degree$ vs. 28$\degree$; both satellites should detect a similar population of cosmic rays. In Section \ref{s:solar_modulation}, we found that the peak-to-peak variation in the observed cosmic-ray particle flux from this work was most consistent with the variation seen in the lowest rigidity bin, $1.01-1.16$ GV, analyzed in \citet{corti2019}. Thus we expect a rigidity derived in this work to be in approximately the same range.

Using the derived estimate of the average kinetic energy for cosmic rays impacting HST of 534 $\pm$ 134 MeV and the relationship in Appendix B, we compute a momentum of 1134.43 MeV/c $\approx$ 1.1 GeV/c. In these units, the rigidity is 1.1 GV which agrees with our expectation and confirms the validity of our approach to estimating the kinetic energy of the average cosmic ray.


\section{Summary}
We developed a Python package, \texttt{HSTCosmicrays}, and used it to characterize transient, high energy particles detected in dark frames that were taken with 5 CCD imagers on HST; ACS/HRC, ACS/WFC, STIS/CCD, WFPC2, and WFC3/UVIS. Whenever possible, the software will perform a connected-component labeling analysis on the DQ array to identify groups of on pixels marked with 8192 (i.e. cosmic ray). When that is not possible, the software will use a technique we refer to as threshold labeling to identify cosmic rays while rejecting hot pixels. For every observation, we record a variety of image metadata and store morphological parameters for each cosmic ray identified. 

In total, we have characterized approximately 1.2 billion cosmic rays. We modeled the energy-loss distributions for 3 of the 5 CCD imagers and estimated that the typical particle observed by HST has a kinetic energy of 534 $\pm$ 117 MeV. We qualitatively reproduced the overall structure observed in the thickness maps derived from fringing anaylses for ACS/HRC, ACS/WFC, and WFC3/UVIS. Next, we observed anti-correlation between solar activity and cosmic-ray particle flux. A spectral analysis of the cosmic-ray particle flux over time revealed two signals with peaks at $\sim$11 years (the solar cycle) and $\sim$48 days (unknown origin). The modulation of the cosmic-ray particle flux by the solar cycle indicates that the majority of the cosmic rays observed by HST are galactic cosmic rays. 

We compiled a variety of useful statistics describing the distributions of the morphological parameters computed for cosmic rays observed in each of the 5 CCD imagers. These values can be used to quantify the impact of cosmic rays on observations with ACS, STIS, and WFC3 for proposal planning purposes. For example, the number of pixels affected by cosmic rays in a given observation can be computed using the cosmic-ray particle flux and the number of pixels a cosmic ray will affect on average. 

We analyzed observations made with the STIS/CCD taken during passages through the center of the SAA. We found that the cosmic-ray particle flux increased by a factor of 800 over a span of $\sim350$ seconds from the edge of the SAA to the center. Using the rate of energy deposition, we estimated the peak cosmic-ray particle flux to be 1092.79 $CR/s/cm^2$ at a location of 41.61 $\degree$ W  and 22.96 $\degree$S which is in good agreement with the contemporaneous centroid derived by \citet{furst2009}. A spatial analysis of 54,215 observations identified two ``hot spots", one over North America and one extending from South Africa to the western coast of Australia, where the cosmic-ray particle flux increases to more than 5$\sigma$ above the nominal value. 

\subsection{Acknowledgements}
The data presented in this paper were obtained from the Mikulski Archive for Space Telescopes (MAST) and its cloud-hosted counterpart, \href{https://registry.opendata.aws/hst/}{HST Public Dataset}, on AWS. STScI is operated by the Association of Universities for Research in Astronomy, Inc., under NASA contract NAS5-26555.  
Support for NDM and SED was provided by NASA through grant number 14587 from the Space Telescope Science Institute, which is operated by AURA, Inc., under NASA contract NAS 5-26555. 
Thank you to Chris Long for  detailed engineering discussions and information on HST operations, and, to Jeremy Walsh and Michael Wong for discussions on fringing models.  We also thank the referee whose comments and suggestions greatly improved this paper.

This research has made use of NASA's Astrophysics Data System.
This work made use of the following open source Python packages; \texttt{astropy}, \texttt{astroquery}, \texttt{cartopy}, \texttt{dask}, \texttt{iminuit}, \texttt{matplotlib}, \texttt{numpy}, \texttt{pandas}, and \texttt{statsmodels}.

\bibliography{master_isr}

\appendix

\section{STIS SAA Observations}

\begin{table}[H]
    \caption{The estimated cosmic-ray particle flux and counts computed for the STIS/CCD SAA dataset. Exposure time for all images was 60 seconds.}
    \hspace{-2cm}
    \begin{tabular}{ccccccc}
\toprule
Position & OBSID &  Latitude &  Longitude & Rate of Energy &  Cosmic Ray  & Number of  \\
&  & & & Deposition & Flux 	 & Cosmic Rays\\
&  & degree    & degree  & $e^-/s/cm^2$ & $CR/s/cm^2$ &  \\
             \midrule
1&  o3st20gcq\_flt.fits &           -3.81 &           -82.76 &                              $3.57\mathrm{x}{10^3}$ &             1.36 &    472.52 \\
2& o3st20gdq\_flt.fits &           -6.96 &           -77.26 &                             $3.68\mathrm{x}{10^4}$ &            14.04 &   4868.80 \\
3& o3st20geq\_flt.fits &          -10.03 &           -71.68 &                            $2.35\mathrm{x}{10^5}$  &            89.59 &  $3.11\mathrm{x}{10^4}$ \\
4& o3st20gfq\_flt.fits &          -12.99 &           -66.00 &                            $5.91\mathrm{x}{10^5}$ &           225.32 &  $7.81\mathrm{x}{10^4}$ \\
5& o3st20ggq\_flt.fits &          -15.80 &           -60.17 &                           $1.22\mathrm{x}{10^6}$ &           466.61 & $1.62\mathrm{x}{10^5}$ \\
6& o3st20ghq\_flt.fits &          -18.43 &           -54.18 &                           $1.92\mathrm{x}{10^6}$ &           730.41 & $2.53\mathrm{x}{10^5}$ \\
7& o3st20giq\_flt.fits &          -20.83 &           -47.99 &                           $2.68\mathrm{x}{10^6}$ &          1021.34 & $3.54\mathrm{x}{10^5}$ \\
8& o3st20gjq\_flt.fits &          -22.96 &           -41.61 &                           $2.86\mathrm{x}{10^6}$ &          1092.79 & $3.79\mathrm{x}{10^5}$ \\
9& o3st20gkq\_flt.fits &          -24.79 &           -35.03 &                           $1.98\mathrm{x}{10^6}$ &           755.76 & $2.62\mathrm{x}{10^5}$ \\
10& o3st20glq\_flt.fits &          -26.28 &           -28.25 &                          $1.12\mathrm{x}{10^6}$ &           428.76 & $1.49\mathrm{x}{10^5}$ \\
11& o3st20gmq\_flt.fits &          -27.41 &           -21.32 &                          $6.08\mathrm{x}{10^5}$ &           231.84 &  $8.04\mathrm{x}{10^4}$ \\
12& o3st20gnq\_flt.fits &          -28.13 &           -14.27 &                            $2.56\mathrm{x}{10^5}$ &            97.66 &  $3.39\mathrm{x}{10^4}$ \\
13& o3st20goq\_flt.fits &          -28.44 &            -7.14 &                            $1.19\mathrm{x}{10^5}$ &            45.52 &  $1.58\mathrm{x}{10^4}$ \\
14& o3st20gpq\_flt.fits &          -28.33 &            -0.00 &                             $5.57\mathrm{x}{10^4}$ &            21.26 &   7373.38 \\
15& o3st20gqq\_flt.fits &          -27.80 &             7.09 &                             $3.48\mathrm{x}{10^4}$ &            13.29 &   4609.25 \\
16& o3st20grq\_flt.fits &          -26.87 &            14.07 &                             $2.61\mathrm{x}{10^4}$ &             9.95 &   3450.12 \\
17& o3st20gsq\_flt.fits &          -25.55 &            20.92 &                             $1.57\mathrm{x}{10^4}$ &             5.97 &   2071.02 \\
18& o3st20gtq\_flt.fits &          -23.88 &            27.58 &                              $5.79\mathrm{x}{10^3}$ &             2.21 &    766.55 \\
19& o3st20guq\_flt.fits &          -21.90 &            34.04 &                              $4.49\mathrm{x}{10^3}$ &             1.71 &    594.29 \\
\bottomrule
\bottomrule
\end{tabular}
    \label{tab:stis_saa_estimates}
\end{table}

\section{Useful Relativistic Relationships}\label{s:relativistic_relationships}
\subsection{Momentum and Kinetic Energy}

The relativistic momentum-energy relation defines a relationship between the total relativistic energy, $E$, the relativistic momentum, $p$, and the rest mass, $m_0$ given by,

\begin{equation}
    E^2 = (pc)^2 + (m_oc^2)^2.
    \label{eq:momentum_relation}
\end{equation}
The total relativistic energy given by $E = \gamma m_o c^2$ and the relativistic momentum is given by $p=\gamma m_o v$. Using Equation \ref{eq:momentum_relation}, we can express the relativistic momentum of a particle in terms of the relativistic kinetic energy, $T = (\gamma - 1)*m_oc^2$, as follows,
\begin{equation}
\begin{split}
    (pc)^2 + (m_o c^2)^2 ={}& (T + m_oc^2)^2  \\
    (pc)^2 ={}& (T + m_o c^2)^2  - (m_o c^2)^2 \\
    p ={}& \frac{\sqrt{T^2 + 2Tm_oc^2}}{c}
\end{split}
\end{equation}

\subsection{Rigidity}
For a relativistic particle traveling in a uniform magnetic field, the Lorentz force will cause the particle to undergo uniform circular motion around the field lines. The magnetic rigidity is the product of the radius of the orbit and the magnetic field strength,
\begin{equation}
    R = B\rho = \frac{pc}{Ze},
    \label{eq:rigidity}
\end{equation}
where $p$ is the momentum, $c$ is the speed of light, $Z$ is the charge number, and $e$ is the elementary charge.
For relativistic particles, the momentum is often quoted in units of GeV/c. When using these units in Equation \ref{eq:rigidity}, the rigidity of a 1 GeV/c particle will be, 

\begin{equation}
\begin{split}
        R = {}& \frac{pc}{Ze} \\
        R = {}& \frac{(1 \mathrm{GeV/c})c}{Ze} \\
        R = {}& \frac{\textrm{$10^9$ eV}}{Ze}\\
        R = {}& \frac{10^9}{Z} \left(\frac{1.602\mathrm{x}10^{-19}
        \mathrm{J}}{e}\right)\\
        R = {}& \frac{10^9}{Z} \left(\frac{1.602\mathrm{x}10^{-19}
        \mathrm{J}}{1.602\mathrm{x}10^{-19} \mathrm{C}}\right)\\
        R = {}& \frac{10^9 \mathrm{V}}{Z} = 1 \mathrm{GV}/Z. 
\end{split}
\end{equation}

Thus we see that for a proton that has a momentum of 10 GeV/c, it has a magnetic rigidity of 10 GV.

\end{document}